\documentclass{article}
\usepackage{graphicx}  % standard LaTeX graphics tool;
\usepackage{amsmath}   % For getting proper math eqns
\usepackage{amssymb}   % Used for getting various symbols eg. gtrsim, lesssim etc.
\usepackage{bm} % For bold math (esp of lower greek letters)
\usepackage{dcolumn}% Align table columns on decimal point
\usepackage{color}
\usepackage{mathrsfs}
\usepackage{amsfonts}
\usepackage{varioref}
\usepackage[utf8]{inputenc}
\usepackage{amsmath}
\usepackage{amsfonts}
\usepackage{amssymb}
\usepackage{enumitem}
\usepackage{float}
\usepackage{epsfig}
\usepackage{amsfonts}
\usepackage{latexsym}
\usepackage{hyperref}
\usepackage{setspace}
\usepackage{bm}
\usepackage{slashed}
\usepackage{graphicx}
\usepackage{bm}
\usepackage{hyperref}
\usepackage{dsfont}
\usepackage{caption}
\usepackage{subcaption}
\def\be{\begin{equation}}
\def\ee{\end{equation}}
\labelformat{section}{Section #1} 
\labelformat{subsection}{Section #1} 
\labelformat{subsubsection}{Section #1}
\labelformat{subsubsubsection}{Section #1}
\labelformat{equation}{Eq.~(#1)} 
\labelformat{figure}{Fig.~#1} 
\labelformat{subfigure}{Fig.~\thefigure#1} 
\labelformat{table}{Tab.~#1} 
\labelformat{appendix}{Appendix #1}
\title{\bf Some measures  for fermionic entanglement in the cosmological de Sitter spacetime }
\author{Sourav Bhattacharya$^1$\footnote{sbhatta@iitrpr.ac.in}, ~~Himanshu Gaur$^{1,2}$\footnote{194123018@iitb.ac.in} ~~and ~Nitin Joshi$^{1}$\footnote{2018phz0014@iitrpr.ac.in}\\
$^1${\small  Department of Physics, Indian Institute of Technology Ropar,  Rupnagar, Punjab 140 001,
India}\\
$^2${{\small  Department of Physics, Indian Institute of Technology Bombay,  Mumbai 400 076,
India\footnote{Current affiliation}}}}

\begin{document}
\maketitle
\begin{abstract}
\noindent We investigate two measures of quantum correlations and entanglement, namely the violation of the Bell-Mermin-Klyshko (BMK) inequalities and the quantum discord, for Dirac fermions in the cosmological de Sitter background of dimension four. The BMK violation is focussed on the vacuum for two and four mode squeezed states and the maximum violation is demonstrated. For the quantum discord, we investigate a maximally entangled in-vacuum state. Qualitative similarities as well as differences of our results with that of  different coordinatisations of de Sitter  in the context of  scalar and fermionic  field theory is discussed.
\end{abstract}
\vskip .2cm
\noindent
{\bf Keywords :}  de Sitter spacetime, fermions, BMK inequality, quantum discord
\bigskip
%\newpage
%\tableofcontents
%%%%%%%%%%%%%%%%%%%%%%%%
\section{Introduction}
%%%%%%%%%%%%
Quantum entanglement is a highly counterintuitive feature of quantum mechanics   related to its non-local characteristics~\cite{Einstein, bell:1964, clauser:1969, cirel:1980, Aspect1, Aspect2, Zyczkowski:1998yd, Vidal:2002zz, chen:2,  mermin:1990, belinski, Gisin:1998, nagata:2002, yu:2003, uffink:2002, Zurek, Dakic, Huang, aharon:2011}. The subject has been put on a firm physical ground since its first experimental verification~\cite{Aspect1, Aspect2} and in recent times it has developed various disciplines involving the quantum information and computation, see~\cite{Nielsen:2000} and references therein.

The relativistic sector, where particle pair creation can take place  is always interesting in this context, for such pairs turn out to be entangled. The most well studied case of quantum entanglement in this context corresponds to the Rindler left-right wedges (or the maximally extended near horizon geometry of a non-extremal black hole)~\cite{Alsing:2003es, alsing:2006, Friis:2011fy, A.dutta, wang, MartinMartinez:2010ar, Brown:2012iz, Qiang, Wang, Yao, Xiao:2018cxg} (also references therein). Due to pair creation, the entanglement or quantum correlation between two maximally entangled Bell pair gets degraded  in the Rindler frame, e.g.~\cite{MartinMartinez:2010ar}. See e.g.~\cite{Bhattacharya:2020sjr} and references therein for a discussion on entanglement in the context of Schwinger pair creation. We further refer our reader to~\cite{Tomaras:2019sjq} and references therein for a discussion on entanglement in the soft sector of quantum electrodynamics.

Another well motivated sector of the relativistic entanglement is the cosmological scenario, important chiefly because such study might provide us insight about the initial state as well as the geometry of the early inflationary universe. Such studies might involve investigations like the vacuum entanglement entropy, various other measures of quantum entanglement between initially entangled states for both bosonic and fermionic fields as well as the quantum decoherence of cosmological perturbations and their possible observational consequences,
e.g.~\cite{Fuentes:2010dt, Maldacena:2012xp, Lim, kanno:2015, Maldacena:2015bha, Vennin, Liu:2016aaf, Kanno:2016gas, Kanno:2017dci, Choudhury:2016cso, Chen:2017cgw, Choudhury:2017bou, Albrecht:2018prr, Feng:2018ebt, Bhattacharya:2018yhm, dePutter:2019xxv, Bhattacharya:2019zno, Matsumura:2020uyg, Brahma:2020zpk} and references therein.

In this work we wish to investigate two measures -- namely the violation of the Bell-Mermin-Klyshko inequality~\cite{bell:1964, clauser:1969,  mermin:1990, belinski, Gisin:1998} and the quantum discord~\cite{Zurek, Dakic} for Dirac fermions in the cosmological de Sitter background.  

The Bell inequality~\cite{bell:1964} (see also~\cite{Nielsen:2000} and references therein) is a measure of non-locality for a two-partite quantum system. Later such inequality was extended to multipartite systems~\cite{clauser:1969,  mermin:1990, belinski, Gisin:1998}, altogether  regarded as  the Bell-Mermin-Klyshko (BMK) inequalities. In the nonlocal regime of quantum mechanics BMK inequalities may be violated, thereby clearly distinguishing quantum effects from that of any local classical hidden variables. As the number of partite is increased in a system,  the upper bound of the BMK violation also increases, e.g.~\cite{nagata:2002, yu:2003}.  Given two subsystems, on the other hand, quantum discord is a suitable measure of all correlations including entanglement between them~\cite{Zurek, Dakic}. Accordingly, even if there is no entanglement for a mixed state, the  quantum discord can be non-vanishing. The key ingredient of the computation of discord is the quantum mutual information between the subsystems. One also needs to optimise over all possible measurements performed on one of the subsystems. 

We refer our reader to~\cite{Huang,  Qiang, Wang, Yao, Xiao:2018cxg, Lim,  Maldacena:2015bha, Vennin, Kanno:2016gas, Kanno:2017dci, Choudhury:2016cso, Chen:2017cgw, Choudhury:2017bou, Feng:2018ebt, Matsumura:2020uyg} and references therein for discussions on the BMK violation and quantum discord in both non-inertial and inflationary scenarios.\\   

The basic computational tools we shall use in this paper can be seen in~\cite{Kanno:2016gas, Kanno:2017dci} and references therein. Our study on the BMK violation will be focused on the vacuum, whereas for the quantum discord we shall work on some maximally entangled initial Bell state, which will give us insight about correlations between non-vacuum states.  In~\cite{Kanno:2016gas}, the quantum discord corresponding to a maximally entangled state for two scalar fields  was investigated in the hyperbolic de Sitter background. In~\cite{Kanno:2017dci}, the infinite BMK violation was demonstrated for a massless scalar field in a cosmological background which is de Sitter and radiation dominated respectively in the past and future. We shall compute these two measures for massive Dirac fermions in the cosmological de Sitter background in order to see how much similar or dissimilar the result is with the already existing results.

The rest of the paper is organised as follows. In the next section we construct the relevant two and four mode squeezed states using the Bogoliubov relations discussed in \ref{A}. We compute the BMK violation for the two and four mode squeezed states in \ref{Bell}. Computation of the discord can be seen in \ref{discord}. Finally we discuss the results and conclude in \ref{con}.

We work in spacetime dimension four and set $c=1=\hbar$.

%%%%%%%%%%%
\section{Fermionic squeezed states} \label{sqz}
%%%%%%%%%%%%%
Based upon the discussion of \ref{A}, we shall construct below the two- and four-mode fermionic squeezed states, to be useful for our purpose. Corresponding to the field quantisations \ref{f1}, \ref{f2}, we define the `in' and `out' vacua as,
$$
a_{\rm in}(\vec{k},s) |0_{\rm in}\rangle=0= b_{\rm in}(\vec{k},s) |0_{\rm in}\rangle,            \qquad {\rm and } \qquad a_{\rm out}(\vec{k},s) |0_{\rm out}\rangle=0= b_{\rm out}(\vec{k},s) |0_{\rm out}\rangle
$$
The Bogoliubov relations of  \ref{bglv3} show that the `in' vacuum can be expressed as a squeezed state over all the `out' states,
\begin{eqnarray}
|0_{\rm in}\rangle \sim \exp\left[-\sum_{\vec{k}, s} \frac{\beta_{k}} {\alpha_{k}^\ast} a^{\dagger}_{\rm out}(\vec{k},s) b^{\dagger}_{\rm out}(-\vec{k},s)     \right] |0_{\vec{k},s,\,{\rm out}}\rangle \otimes|0_{-\vec{k},s,\,{\rm out}}\rangle,
\label{sq1}
\end{eqnarray}
where $|0_{\pm \vec{k},s,\,{\rm out}}\rangle$ respectively represent a particle and an antiparticle vacuum. 

We shall work here with a specific value of the spatial momentum, $\vec{k}$, e.g.~\cite{Kanno:2016gas}.
We also note from \ref{bglv3} that the helicities do not mix in the Bogoliubov transformations. Thus due to the various anti-commutation relations, the squeezed state expansion corresponding to different $s$ values in \ref{sq1} will just factor out.  This permits us to go for another simplification -- to restrict ourselves to a specific $s$ value as well. In other words, we shall work with a subspace of $|0_{\rm in}\rangle$ corresponding to specific $\vec{k}$ and $s$. Thus instead of \ref{sq1}, we work with (after normalisation)
\begin{eqnarray}
|0_{{\vec{k}}, \rm in}\rangle= |\alpha_k| \left[ |0_{\vec{k},\,{\rm out}}\rangle \otimes|0_{-\vec{k},\,{\rm out}}\rangle - \frac{\beta_{k}} {\alpha_{k}^\ast} |1_{\vec{k},\,{\rm out}}\rangle \otimes|1_{-\vec{k},\,{\rm out}}\rangle\right],
\label{sq2}
\end{eqnarray}
where we have suppressed the index $s$, since we are restricting ourselves to   any single value of it. We shall further comment on the more general helicity summed state at the end of \ref{Bell2}.  The above is called a two-mode squeezed state.\\

\noindent
The notion of the two-mode squeezed state can easily be extended if we include more than one fermionic fields, say $\Psi_1(x), \, \Psi_2(x), \dots$, each quantised in a way described in \ref{A} and further mix these particle species via some Bogoliubov transformations (see \cite{Kanno:2017dci} for discussions on scalar field theory, also~\cite{Blasone, Blasone2}).

Let us consider two fermionic fields, $\Psi_1$ and $\Psi_2$ with their `in' vacuum $|0_{{\rm in}}\rangle_1$ and $|0_{{\rm in}}\rangle_2$ respectively,
\begin{eqnarray}
a_{{\rm in},i}(\vec{k},s) |0_{{\rm in}}\rangle_i=0= b_{{\rm in}, i}(\vec{k},s) |0_{{\rm in}}\rangle_i \qquad (i=1,2,\,\,\,{\rm no~sum~on}~i),           
\label{sq3}
\end{eqnarray}
where $a_{{\rm in},1},\,b_{{\rm in},1}$ and $a_{{\rm in},2},\,b_{{\rm in},2}$ are the annihilation operators corresponding to $\Psi_1$ and $\Psi_2$ respectively.
The combined  `in' vacuum for these two field system is then $|0_{{\rm in}}\rangle_1 \otimes |0_{{\rm in}}\rangle_2  $. 
Let us suppose that these two fields are correlated via a simple mixing transformation as,
\begin{eqnarray}
\begin{split}
\overline{a}^{(1)}_{{\rm in}}(\vec{k},s)=\Gamma_{k} \,a_{{\rm in},1}(\vec{k},s)+\Delta_{k}\,b^{\dagger}_{{\rm in}, 2}(-\vec{k},s), \qquad  \overline{a}^{(2)}_{{\rm in}}(\vec{k},s)=\Gamma_{k}\, a_{{\rm in},2}(\vec{k},s)+\Delta_{k}\,b^{\dagger}_{{\rm in}, 1}(-\vec{k},s)  \\
\overline{b}^{(1)}_{{\rm in}}(\vec{k},s)=\Gamma_{k} \,b_{{\rm in},1}(\vec{k},s)-\Delta_{k}\,a^{\dagger}_{{\rm in}, 2}(-\vec{k},s), \qquad  \overline{b}^{(2)}_{{\rm in}}(\vec{k},s)=\Gamma_{k}\, b_{{\rm in},2}(\vec{k},s)-\Delta_{k}\,a^{\dagger}_{{\rm in}, 1}(-\vec{k},s)
\label{sq4}
\end{split}
\end{eqnarray}
where $\Gamma_{k}$ and $\Delta_{k}$ are Bogoliubov coefficients satisfying,  $|\Gamma_{k}|^2+|\Delta_{k}|^2=1$. It is easy to check that the operators defined above  satisfy the canonical anti-commutation relations. We assume that such squeezing  between different field species is weak, i.e.,
$$\left\vert\frac{\Delta_{k}} {\Gamma_{k}}\right \vert \ll 1$$
Let us denote the vacuum state corresponding to the new operators in \ref{sq4} by $|\overline{0}\rangle$, 
\begin{eqnarray}
\overline{a}^{(i)}_{{\rm in}}(\vec{k},s)|\overline{0}\rangle=0=\overline{b}^{(i)}_{{\rm in}}(\vec{k},s)|\overline{0}\rangle \qquad (i=1,2)
\label{sq5}
\end{eqnarray}
From \ref{sq4}, $|\overline{0}\rangle $ can then be expanded as,
\begin {eqnarray}
|\overline{0}\rangle \sim e^{-\sum_{\vec{k}, s} \frac{\Delta_{k}} {\Gamma_{k}^\ast} \left[a^{\dagger}_{{\rm in},1}(\vec{k},s)b^{\dagger}_{{\rm in},2}(-\vec{k},s)+a^{\dagger}_{{\rm in},2}(\vec{k},s)b^{\dagger}_{{\rm in},1}(-\vec{k},s)\right]} |0_{{\rm in}}\rangle_1 \otimes |0_{{\rm in}}\rangle_2 %\nonumber\\
\label{sq6}
\end{eqnarray}
We focus as earlier on a specific $\vec{k}$ and $s$ value. Suppressing the index $s$, making the expansion of the exponential   only up to the second order owing to the weak squuezing, and after normalising, \ref{sq6} takes the form
\begin{eqnarray}
|\overline{0}_{\vec k}\rangle=A_{k}|0_{{\rm in},\,\vec{k}}\rangle_1 \otimes |0_{{\rm in},\, \vec{k}}\rangle_2  +\frac{B_{k}}{\sqrt2}\left(|1_{{\rm in},\,\vec{k}}\rangle_1  \otimes |1_{{\rm in},\,-\vec{k}}\rangle_2+|1_{{\rm in},\,-\vec{k}}\rangle_1  \otimes |1_{{\rm in},\,\vec{k}}\rangle_2\right),
\label{sq7}
\end{eqnarray}
where $A_k$ and $B_k$  depend upon $\Gamma_k$ and $\Delta_k$ and   $|A_{k}|^2+|B_{k}|^2=1$.

The `in' states appearing on the right hand side of \ref{sq7} can further be expanded according to \ref{sq2}. If we take the rest mass of both the fields to be the same, the Bogoliubov coefficients ($\alpha_k,\,\beta_k$) corresponding to these two fields are then same as well, cf. \ref{A}. This yields,
\begin{eqnarray}
|\overline{0}_{\vec k}\rangle=&&A_{k}{|\alpha_k|^2 \left[ |0_{\vec{k},\,{\rm out}}\rangle_1 \otimes|0_{-\vec{k},\,{\rm out}}\rangle_1 - \frac{\beta_{k}} {\alpha_{k}^\ast} |1_{\vec{k},\,{\rm out}}\rangle_1 \otimes|1_{-\vec{k},\,{\rm out}}\rangle_1\right] \otimes  \left[ |0_{\vec{k},\,{\rm out}}\rangle_2 \otimes|0_{-\vec{k},\,{\rm out}}\rangle_{2} - \frac{\beta_{k}} {\alpha_{k}^\ast} |1_{\vec{k},\,{\rm out}}\rangle_2 \otimes|1_{-\vec{k},\,{\rm out}}\rangle_2\right]}\nonumber \\
&&+\frac{B_{k} |\alpha_k|^2}{\sqrt{2} (\alpha_k^{\ast})^2 }\left[{|1_{{\vec{k}},\,{\rm out}}\rangle_1\otimes|0_{{-\vec{k}},\,{\rm out}}\rangle_1}\otimes{|0_{{\vec{k}},\,{\rm out}}\rangle_2\otimes|1_{-{\vec{k}},\,{\rm out}}\rangle_2}+{|0_{{\vec{k}},\,{\rm out}}\rangle_1\otimes|1_{-{\vec{k}},\,{\rm out}}\rangle_1}\otimes{|1_{{\vec{k}},\,{\rm out}}\rangle_2\otimes|0_{-{\vec{k}},\,{\rm out}}\rangle_2}\right], %\nonumber\\
\label{sq8}
\end{eqnarray}
known as the four mode squeezed state.
Note that the above construction is similar but not exactly the same as the de Sitter $\alpha$-vacua (e.g.~\cite{Collins:2004wj}). This is because the latter mixes the modes of a single quantum field whereas here we have mixed two different fields themselves. It is also clear that the construction of such squeezed states goes beyond two fields. For example, with three fermionic fields one can construct a six mode squeezed state. 

Having constructed the necessary states, we are now ready to go into computing the BMK violation and the quantum discord.

%%%%%%%%%
\section{The violation of the  BMK inequalities}\label{Bell}
%%%%%%%%%%

\subsection{The BMK inequalities}\label{Bell1}
	%%%%%%%%%%%%%%
	The construction of the Bell and the Bell-Mermin-Klyshko (BMK) operators for fermions is similar to that of the scalar field theory~\cite{Kanno:2017dci} (also references therein).
Let us consider two sets of non-commuting observables defined respectively over the Hilbert spaces $\mathscr{H}_{X}$ and  $\mathscr{H}_{Y}$ :  $\{X, X^{\prime} \in \mathscr{H}_{X}\}$ and $\{Y, Y^{\prime} \in \mathscr{H}_{Y} \}$. We assume that these are spin-1/2 operators along  specific directions, such as $X= n_i \sigma_i$, $X^{\prime}= n'_i \sigma_i$, where $\sigma_i$'s are the Pauli matrices and $n_i$, $n'_i$ are unit vectors on the three dimensional Euclidean space. The eigenvalues of these operators are $\pm1$. The Bell operator ${\cal B} \in \mathscr{H}_{X}\otimes \mathscr{H}_{Y}$ is defined as
\begin{eqnarray}
{\cal B}=\frac12\left(X \otimes Y+X^{\prime} \otimes Y +X \otimes Y^{\prime} -X^{\prime} \otimes Y^{\prime}\right)=\frac{1}{2}X\otimes(Y+{Y^\prime})+\frac{1}{2}{{X^\prime}}\otimes(Y-{Y^\prime})
\label{b1}
\end{eqnarray}
In theories with local classical hidden variables we have  the so called Bell's inequality, $\langle{\cal B}^2\rangle\leq 1$ and $|\langle{\cal B}\rangle|\leq 1$~\cite{clauser:1969}.  However, this is violated in quantum mechanics. Indeed, from \ref{b1} we have (suppressing the tensor product sign),
\begin{eqnarray}
{\cal B}^2={\bf I} - \frac{1}{4} [X,X^\prime]\,[Y,Y^\prime],
\label{b2}
\end{eqnarray}
where ${\bf I}$ is the identity operator. Using the commutation relations for the Pauli matrices, one gets $\langle{\cal B}^2\rangle\leq2$ or $|\langle{\cal B}\rangle|\leq\sqrt2$, obtaining a violation of Bell's inequality, where the upper bound is known as the {\it maximum violation}~\cite{cirel:1980}.

The above construction can be extended to multipartite systems, known as the Mermin-Klyshko or collectively the Bell-Mermin-Klyshko inequalities.  The relevant operator is recursively defined as,
\begin{eqnarray}
{\cal B}_{n}=\frac{1}{2}{ \cal B}_{n-1}({ \cal O}_{n}+{\cal O}_{n}^\prime)+\frac{1}{2}{\cal B}_{n-1}^\prime({\cal O}_{n}-{\cal O}_{n}^\prime), \qquad n=2,3,4,\dots,
\label{b3}
\end{eqnarray}
where we have defined ${\cal B}_{1} ={\cal O}_{1}$, ${\cal B}^{\prime}_{1} ={\cal O}_{1}^{\prime}$. All the operators correspond to spin-1/2 systems as earlier. 

In classical hidden variable theories we have ${\cal O}_{n}=\pm  {\cal O}_{n}^\prime$, yielding  the  Mermin-Klyshko inequalities
\begin{eqnarray}
|\langle{\cal B}_{n}\rangle|\leq 1, \qquad n=1,2,3, \dots 
\end{eqnarray}
whereas in quantum mechanics one has~\cite{Gisin:1998, nagata:2002, yu:2003},
\begin{eqnarray}
|\langle{\cal B}_{n}\rangle|\leq 2^{\frac{n-1}{2}}, \qquad n=1,2,3,\dots
\label{b4}
\end{eqnarray}
    Thus the BMK inequality will be violated for multipartite states, $n\geq 2$. We shall mention another relevant form of the BMK inequality at the end of the next subsection.

%%%%
\subsection{Computing the  violation } \label{Bell2}
%%%%

\begin{figure}[h!]
%\begin{center}
  \includegraphics[width=8cm]{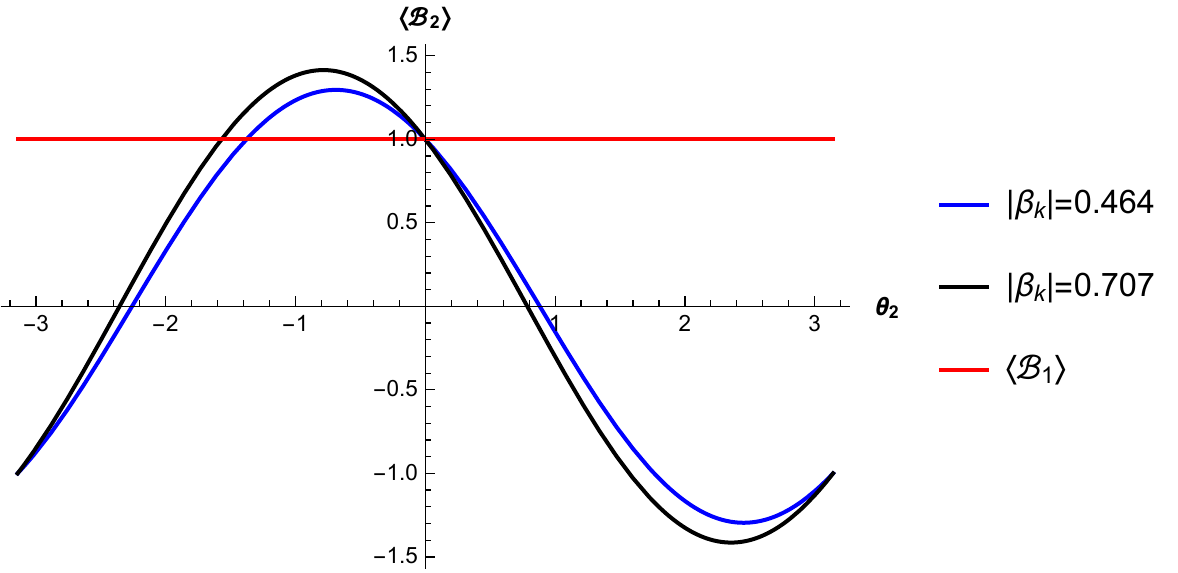}
  \includegraphics[width=8cm]{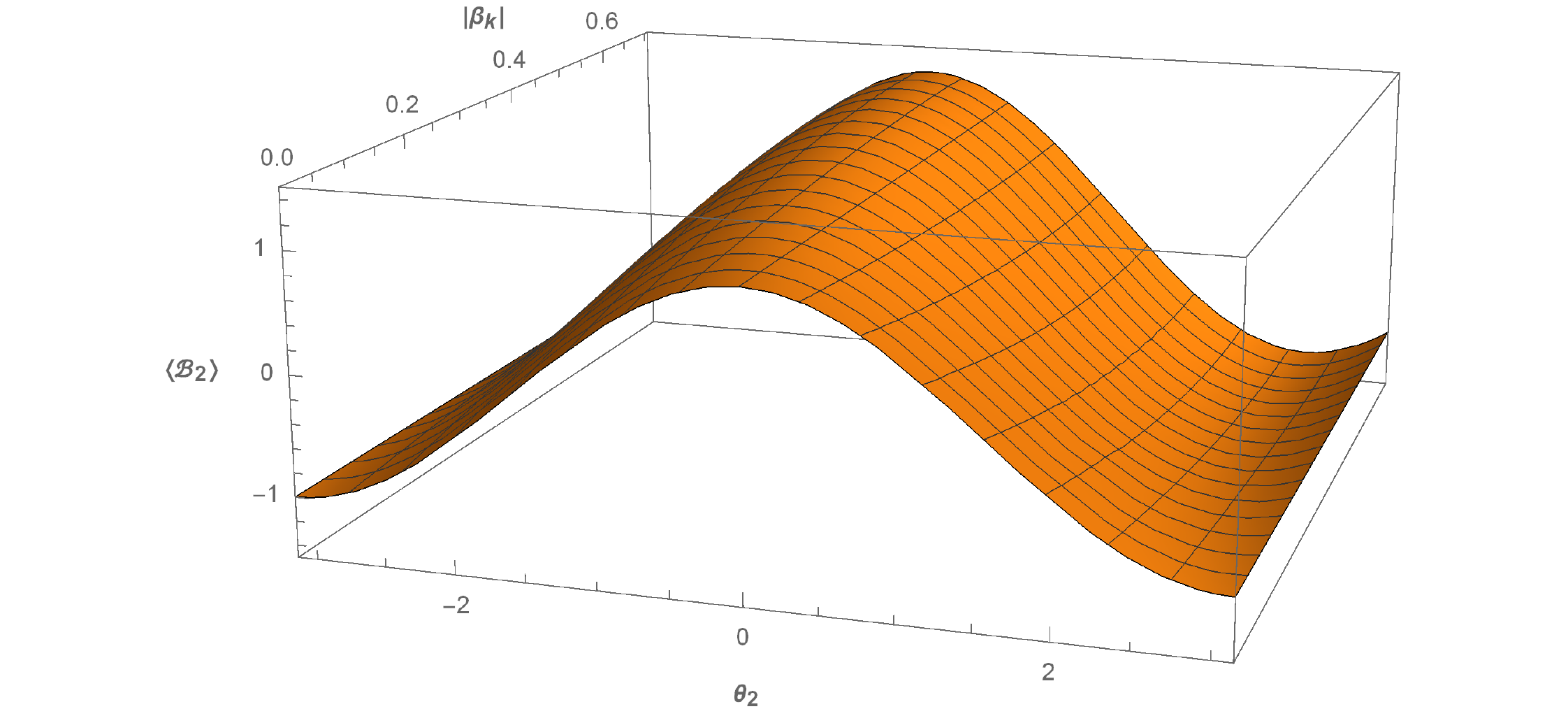}
  \caption{2-d and 3-d plots for the violation of BMK inequalities, \ref{bv3}, for two-mode squeezed state with different values of the Bogoliubov coefficients.   In the 2-d plot, the red line, $\langle{\cal B}_{1}\rangle$, stands for the classical upper bound. $\langle{\cal B}_{2}\rangle \sim \sqrt{2}$ indicates the maximum violation. Note also from \ref{bglv3} that the variation of the Bogoliubov coefficient correspond to the variation of the rest mass  only. Thus for a given particle species, the value of the Bell violation does not change with respect to the momentum $\vec{k}$.   }
  \label{fig:bell1}
\end{figure}

Let us first  compute  the BMK violation for the two-mode squeezed state defined in~\ref{sqz}. Following \cite{chen:2, Kanno:2017dci}, we introduce a pseudospin operator $S$, 
\begin{eqnarray}
\textbf{n}.\textbf{S}:=S_{z}\cos{\theta}+\sin{\theta}(e^{i\phi}S_{-}+e^{-i\phi}S_{+})
\label{bv0}
\end{eqnarray} 
where $\textbf{n}\equiv (\sin{\theta} \cos{\phi},\sin{\theta}\sin{\phi},\cos{\theta})$ is a  spatial unit vector, $S_{\pm}$ are the ladder operators and $\textbf{n}.\textbf{S}$ has eigenvalues $\pm1$. We have the operations over the orthonormal states $|0\rangle$, $|1\rangle$ (corresponding to a fix value of the helicity $s$), 
\begin{eqnarray}
\begin{split}
&S_{z}|0\rangle=-|0\rangle,\,\, S_{z}|1\rangle=|1\rangle; \qquad 
S_{+}|0\rangle=|1\rangle,\,\, S_{+}|1\rangle=0; \qquad
S_{-}|0\rangle=0,\,\, S_{-}|1\rangle=|0\rangle
\label{bv1}
\end{split}
\end{eqnarray}
Since \ref{bv0} is defined on an Euclidean plane, we can use its rotational invariance to set  $\phi=0$ in \ref{bv0}, so that the expectation value of the Bell operator, \ref{b3}, in the two-mode squeezed state 
%
%\begin{eqnarray}
%\textbf{n}\cdot \textbf{S}=S_{z}\cos{\theta}+S_{x}\sin{\theta}.
%\end{eqnarray}  
%
(\ref{sq2}) becomes
\begin{eqnarray}
\langle0_{\vec{k},{\rm in}}|{\cal B}_{2}|0_{\vec{k},{\rm in}}\rangle=\frac{1}{2}[F(\theta_{1},\theta_{2})+F(\theta_{1},\theta_{2}^{\prime})+F(\theta_{1}^{\prime},\theta_{2})-F(\theta_{1}^{\prime},\theta_{2}^{\prime})]
\label{exp}
\end{eqnarray}
where ${\cal O}_{1}=\textbf{n}_{1}.\textbf{S}$,  ${\cal O}_{2}=\textbf{n}_{2}.\textbf{S}$, ${\cal O}_{1}^\prime=\textbf{n}_{1}^\prime.\textbf{S}$, ${\cal O}_{2}^\prime=\textbf{n}_{2}^\prime.\textbf{S}$, in \ref{b3} and $F(\theta_{1},\theta_{2})$ is given by
\begin{eqnarray}
\begin{split} 
F(\theta_{1},\theta_{2})=&\langle0_{\vec{k},{\rm in}}| (S_{z}\cos{\theta_{1}}+S_{x}\sin{\theta_{1}})\otimes(S_{z}\cos{\theta_{2}}+S_{x}\sin{\theta_{2}})|0_{\vec{k},{\rm in}}\rangle ,\\&
=(\cos{\theta_{1}}\cos{\theta_{2}}- 2|\beta_{{k}} \alpha_{{k}}| \sin{\theta_{1}\sin{\theta_{2}}})
\end{split}
\end{eqnarray}
The other $F$'s can be found by replacing  $\theta_1$, $\theta_2$ suitably above. The violation of the BMK inequality can be examined for various values of the angles as well as the Bogoliubov coefficients. 
For example for $\theta_{1}=0$, $\theta_{1}^{\prime}=\pi/2$ and $\theta_{2}=-\theta_{2}^{\prime}$, we obtain
\begin{eqnarray}
\langle0_{\vec{k},{\rm in}}|{\cal B}_{2}|0_{\vec{k},{\rm in}}\rangle=\cos{\theta_{2}}-2|\beta_{{k}} \alpha_{{k}}|\sin{\theta_{2}},
\label{bv3}
\end{eqnarray}
which is plotted in \ref{fig:bell1}. In particular, the maximum violation  $ \langle{\cal B}_{2}\rangle \sim \sqrt{2} $, is achieved for the maximum value of the Bogoliubov coefficient, $|\beta_k| \sim 0.707 $ (cf., \ref{A}). 

\noindent
As \ref{b4} indicates,  the upper bound of the Bell violation can be increased  by going to four or higher mode squeezed states. We now wish to demonstrate such BMK violation for a four-mode squeezed state discussed in \ref{sqz}. Using \ref{b3}, the relevant BMK operator can be written as
\begin{eqnarray}
\begin{split}
{\cal B}_{4}=\frac{1}{4}[&-{\cal O}_{1}\otimes {\cal O}_{2}\otimes {\cal O}_{3}\otimes {\cal O}_{4}-{\cal O}_{1}^\prime \otimes {\cal O}_{2}^\prime \otimes {\cal O}_{3}^\prime \otimes {\cal O}_{4}^\prime +{\cal O}_{1} \otimes {\cal O}_{2} \otimes {\cal O}_{3} \otimes {\cal O}_{4}^\prime
+{\cal O}_{1} \otimes {\cal O}_{2} \otimes {\cal O}_{3}^\prime \otimes {\cal O}_{4}\\&
+{\cal O}_{1} \otimes {\cal O}_{2}^\prime \otimes {\cal O}_{3} \otimes {\cal O}_{4}+{\cal O}_{1}^\prime \otimes {\cal O}_{2} \otimes {\cal O}_{3} \otimes {\cal O}_{4}+{\cal O}_{1} \otimes {\cal O}_{2} \otimes {\cal O}_{3}^\prime \otimes {\cal O}_{4}^\prime+{\cal O}_{1} \otimes {\cal O}_{2}^\prime \otimes {\cal O}_{3} \otimes {\cal O}_{4}^\prime\\&
+{\cal O}_{1}^\prime \otimes {\cal O}_{2} \otimes {\cal O}_{3} \otimes {\cal O}_{4}^\prime
+{\cal O}_{1} \otimes {\cal O}_{2}^\prime \otimes {\cal O}_{3}^\prime \otimes {\cal O}_{4}+{\cal O}_{1}^\prime \otimes {\cal O}_{2} \otimes {\cal O}_{3}^\prime \otimes {\cal O}_{4}+{\cal O}_{1}^\prime \otimes {\cal O}_{2}^\prime \otimes {\cal O}_{3} \otimes {\cal O}_{4}\\&-{\cal O}_{1} \otimes {\cal O}_{2}^\prime \otimes {\cal O}_{3}^\prime \otimes {\cal O}_{4}^\prime
-{\cal O}_{1}^\prime \otimes {\cal O}_{2} \otimes {\cal O}_{3}^\prime \otimes {\cal O}_{4}^\prime-{\cal O}_{1}^\prime \otimes {\cal O}_{2}^\prime \otimes {\cal O}_{3} \otimes {\cal O}_{4}^\prime-{\cal O}_{1}^\prime \otimes {\cal O}_{2}^\prime \otimes {\cal O}_{3}^\prime \otimes {\cal O}_{4}],
\end{split}
\label{fm1}
\end{eqnarray} 
    where ${\cal O}_{i}=n_{i}\cdot\textbf{S}$ and ${\cal O}_{i}^{\prime}=n^{\prime}_{i}\cdot\textbf{S}$, for $i=1,2,3,4$ and $n_i$'s are unit spacelike vectors on the Euclidean $3$-plane as earlier. We shall compute the expectation value of the above operator in the state~\ref{sq8}. Denoting  the first operator appearing within the square bracket on the right hand side of the above equation by $\mathscr{E}_{4}$, we find
\begin{eqnarray}
\begin{split}
    \langle \overline0_{\vec k}|\mathscr{E}_{4}|\overline0_{\vec k}\rangle=&|A_{k}|^2  {_1}\langle0_{\vec{k},{\rm in}}|\otimes  {_2}\langle0_{\vec{k},{\rm in}}|\mathscr{E}_{4}| 0_{\vec{k},{\rm in}}\rangle_{1}\otimes|0_{\vec{k},{\rm in}}\rangle_{2}\\&+ \frac{A_{k}^\ast B_{k}}{\sqrt2}  {_1}\langle0_{\vec{k},{\rm in}}|\otimes  {_2}\langle0_{\vec{k},{\rm in}}|\mathscr{E}_{4}\left(|1_{\vec{k},{\rm in}}\rangle_1 \otimes |1_{-\vec{k},{\rm in}}\rangle_2+|1_{-\vec{k},{\rm in}}\rangle_1 \otimes |1_{\vec{k},{\rm in}}\rangle_2\right)\\&+ \frac{A_{k} B_{k}^\ast}{\sqrt2} \left(_1\langle1_{\vec{k},{\rm in}}| \otimes {_2}\langle1_{-\vec{k},{\rm in}}|+_1\langle1_{-\vec{k},{\rm in}}| \otimes {_2}\langle1_{\vec{k},{\rm in}}|\right)\mathscr{E}_{4}| 0_{\vec{k},{\rm in}}\rangle_{1}\otimes|0_{\vec{k}}^{\rm in}\rangle_{2}\\&+\frac{|B_{k}|^2}{2}\left(_1\langle1_{\vec{k},{\rm in}}| \otimes {_2}\langle1_{-\vec{k},{\rm in}}|+_1\langle1_{-\vec{k},{\rm in}}| \otimes {_2}\langle1_{\vec{k},{\rm in}}|\right)\mathscr{E}_{4}\left(|1_{\vec{k},{\rm in}}\rangle_1 \otimes |1_{-\vec{k},{\rm in}}\rangle_2+|1_{-\vec{k},{\rm in}}\rangle_1 \otimes |1_{\vec{k},{\rm in}}\rangle_2\right)\\&=F(\theta_{1},\theta_{2}, \theta_{3}, \theta_{4}, \phi_{1}, \phi_{2}, \phi_{3}, \phi_{4})~~~({\rm say}).
\end{split}
\label{f1'}
\end{eqnarray}
We obtain after some algebra,
\begin{eqnarray}
\begin{split}
F(\theta_{1},\theta_{2}, \theta_{3}, \theta_{4}, \phi_{1}, \phi_{2}, \phi_{3}, \phi_{4})&=|A_{k}|^2 f(\theta_{1},\theta_{2}, \phi_{1}, \phi_{2})  f(\theta_{3},\theta_{4}, \phi_{3}, \phi_{4})\\&+ \frac{A_{k}^\ast B_{k}}{\sqrt2}[ g_{+}(\theta_{1},\theta_{2}, \phi_{1}, \phi_{2}) g_{-}(\theta_{3},\theta_{4}, \phi_{3}, \phi_{4})+g_{-}(\theta_{1},\theta_{2}, \phi_{1}, \phi_{2}) g_{+}(\theta_{3},\theta_{4}, \phi_{3}, \phi_{4})]\\&+ \frac{A_{k} B_{k}^\ast}{\sqrt2}[ g_{+}^\ast (\theta_{1},\theta_{2}, \phi_{1}, \phi_{2}) g_{-}^\ast (\theta_{3},\theta_{4}, \phi_{3}, \phi_{4})+g_{-}^\ast (\theta_{1},\theta_{2}, \phi_{1}, \phi_{2}) g_{+}^\ast (\theta_{3},\theta_{4}, \phi_{3}, \phi_{4})]\\&+\frac{|B_{k}|^2}{2}[ h_{++} (\theta_{1},\theta_{2}, \phi_{1}, \phi_{2}) h_{--} (\theta_{3},\theta_{4}, \phi_{3}, \phi_{4})+h_{--} (\theta_{1},\theta_{2}, \phi_{1}, \phi_{2}) h_{++} (\theta_{3},\theta_{4}, \phi_{3}, \phi_{4})\\&+ h_{+-} (\theta_{1},\theta_{2}, \phi_{1}, \phi_{2}) h_{-+} (\theta_{3},\theta_{4}, \phi_{3}, \phi_{4})+h_{-+} (\theta_{1},\theta_{2}, \phi_{1}, \phi_{2}) h_{+-} (\theta_{3},\theta_{4}, \phi_{3}, \phi_{4})]
\end{split}
\label{f2'}
\end{eqnarray}
where we have denoted for the sake of brevity,
\begin{eqnarray}
%\begin{split}
 &&f(\theta_{1},\theta_{2},\phi_{1},\phi_{2})= {_1}\langle0_{\vec{k},{\rm in}}|(\textbf{n}_{1}.\textbf{S})\otimes (\textbf{n}_{2}.\textbf{S})| 0_{\vec{k},{\rm in}}\rangle_{1}=(\cos{\theta_{1}} \cos{\theta_{2}}- 2|\beta_{k} \alpha_{k}| \cos(\phi_{1}+\phi_{2}) \sin{\theta_{1}}\sin{\theta_{2}}) \nonumber\\
&& g_{+}(\theta_{1},\theta_{2},\phi_{1},\phi_{2})={_1}\langle0_{\vec{k},{\rm in}}|(\textbf{n}_{1}.\textbf{S})\otimes (\textbf{n}_{2}.\textbf{S})| 1_{\vec{k},{\rm in}}\rangle_{1}
=\frac{|\alpha_k|^2}{\alpha_k^\ast} \left(-\sin{\theta_{1}} \cos{\theta_{2}}e^{i \phi_{1}} - \left\vert \frac{{\beta_{k}}}{ {\alpha_{k}}}\right \vert e^{-i\phi_{2}} \cos{\theta_{1}}\sin{\theta_{2}}\right)  \nonumber\\
 &&g_{-}(\theta_{1},\theta_{2},\phi_{1},\phi_{2})={_1}\langle0_{\vec{k},{\rm in}}|(\textbf{n}_{1}.\textbf{S})\otimes (\textbf{n}_{2}.\textbf{S})| 1_{-\vec{k},{\rm in}}\rangle_{1}
=\frac{|\alpha_k|^2}{\alpha_k^\ast} \left(-\sin{\theta_{2}} \cos{\theta_{1}}e^{i\phi_{2}} - \left\vert \frac{{\beta_{k}}}{ {\alpha_{k}}}\right \vert  e^{-i\phi_{1}}\cos{\theta_{2}}\sin{\theta_{1}}\right)   \nonumber\\
&& h_{++}(\theta_{1},\theta_{2},\phi_{1},\phi_{2})={_1}\langle1_{\vec{k},{\rm in}}|(\textbf{n}_{1}.\textbf{S})\otimes (\textbf{n}_{2}.\textbf{S})| 1_{\vec{k},{\rm in}}\rangle_{1}
= -\cos{\theta_{1}}\cos{\theta_{2}}= h_{--}(\theta_{1},\theta_{2},\phi_{1},\phi_{2})\nonumber\\
&&h_{+-}(\theta_{1},\theta_{2},\phi_{1},\phi_{2})={_1}\langle1_{\vec{k},{\rm in}}|(\textbf{n}_{1}.\textbf{S})\otimes (\textbf{n}_{2}.\textbf{S})| 1_{-\vec{k},{\rm in}}\rangle_{1}
=\sin{\theta_{1}}\sin{\theta_{2}}e^{-i(\phi_{1}-\phi_{2})} 
=h_{-+}(\theta_{1},\theta_{2},\phi_{1},\phi_{2})^\ast
%\end{split}
\label{b4'}
\end{eqnarray}
The  expectation values corresponding to the other operators in \ref{fm1} can be found by suitably permuting their angular arguments in \ref{f1'}, \ref{f2'} and \ref{b4'}. 

\begin{figure}[h!]
\begin{center}
  \includegraphics[width=8cm]{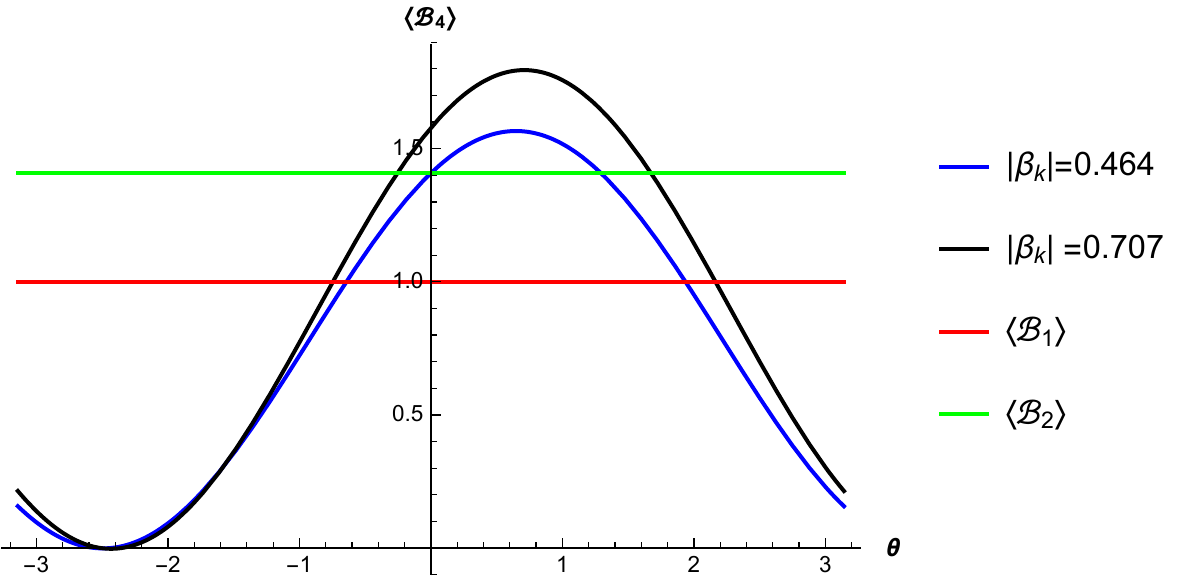}
  \caption{Variation of $\langle{\cal B}_{4}\rangle$ (c.f. \ref{fm1}), with respect to the angular variable $\theta$ and the  subsequent violation of the BMK inequality for four-mode squeezed state. The red line denotes the classical upper bound whereas the green line stands for the quantum upper bound for the two-mode squeezed state. The blue and black lines correspond to  two different values of $|\beta_{k}|$ with the fixed values of $A_{k}\sim 0.975$ and $B_{k}\sim 0.224$ in \ref{sq8}. Thus the maximum violation is larger here compared to the two-mode squeezed state, \ref{fig:bell1}.}
  \label{fig:bell3}
\end{center}
\end{figure}

\ref{fig:bell3} depicts the  expectation value of \ref{fm1} with respect to the angular variable $\theta$.\footnote{We have set all $\phi$'s to zero in \ref{b4'}. Also we have taken, $\theta_1= -0.31^{\circ},\,
\theta_2=40.8^{\circ},\,
\theta_3=-37.1^{\circ},\,
\theta_4=-37.3^{\circ},$
$\theta'_1=-90.5^{\circ},\,
\theta'_2=-40.4^{\circ},\,
\theta'_3=36.9^{\circ},\,
\theta'_4=37.1^{\circ}$.}
 We have taken $A_{k}\sim 0.975$ and $B_{k}\sim 0.224$ in the expression of the four-mode squeezed state, \ref{sq8}. Comparison with \ref{fig:bell1} shows that the BMK violation has increased, in agreement with \ref{b4}. 

Let us now recall that for a general many particle  quantum state which can be grouped into entangled and non-entangled parts, the  BMK violation  can also  be written in a more illuminating form~\cite{uffink:2002},
\begin{eqnarray}
\langle{\cal B}_{N}\rangle^2+\langle{\cal B}^{\prime}_{N}\rangle^2 \leq 2^Z
\label{b6}
\end{eqnarray}
where $N$ represents the total number of partite states and 
\begin{eqnarray}
Z=N-K_{1}-2L+1,
\label{b7}
\end{eqnarray}
where $K_{1}$ denotes the number of single partite states which are not entangled with other $(N-1)$-partite states. $L=\sum_{l=2}^M K_{l}$,  with $K_{l}$ being the number of groups consisting of ${l}$ entangled partite states and $M$ stands for the largest number of such states in a group.  It follows that $N=\sum_{l=1}^M lK_{l}$. \ref{b6} indicates that the upper bound of the BMK violation increases with the number of modes in the squeezed states discussed in \ref{sqz} as follows. For a two-mode squeezed state, we have $N=2,\,L=1,\, K_{1}=0$, so that $Z=1$. For a four mode squeezed state on the other hand, we have $N=4$, $L=1$, $K_1=0$ and hence $Z=3$. Following~\cite{Kanno:2017dci} as of the scalar field, it is then easy to argue that if we include more than one  $\vec{k}$ value into the state, for the two-mode squeezed state the BMK violation does not increase any further, but for the four mode squeezed state it increases without bound, eventually leading to an infinite violation of the BMK inequality. We shall not go into any detail of this.

Finally, we note that we have worked with states with a single value of the helicity  $s$,  \ref{sq2}, \ref{sq8}. It is easy to see that the above results will remain unchanged even if we work with a squeezed state where the $s$ values are summed over, as follows. As we have argued below \ref{sq1}, any squeezed state expansion that sums over  $s$, will factor out between two normalised sub-sectors corresponding to those $s$ values. On the other hand, since  the pseudospin operator  \ref{bv0} acts on states only with a specific $s$ value, it is clear that the  bra and ket for the squeezed state expansion corresponding to the other $s$ value will just give unity, while computing the expectation values like \ref{exp}.  

Finally, we note the qualitative similarity between the BMK violations for a scalar field~\cite{Kanno:2017dci} with that of the Dirac fermions. The BMK violation discussed above basically probes the vacuum state of the cosmological de Sitter spacetime. We also wish to discuss in this work some correlation properties (both classical and quantum) associated with the maximally entangled states. The quantum discord is one such suitable measure, which we investigate below using the two-mode squeezed state, \ref{sq2}.

%%%%%%%%%%%%
\section{Quantum discord}\label{discord}
%%%%%%%%%%%%%

Let us first define the quantum discord~\cite{Zurek, Dakic}.
In classical information theory we have the mutual information between two random variables $X$ and $Y$, defined  as
\begin{eqnarray}
{\cal I}(X,Y)=H(X)+H(Y)-H(X,Y),
\label{d1}
\end{eqnarray}
where 
$$H(X)=-\sum_{X} P(X)\log P(X),  \qquad H(Y)=-\sum_{Y} P(Y)\log  P(Y))$$
 are the Shannon entropies with respective probabilities $P(X)$ and $P(Y)$, and 
 $$H(X,Y)=-\sum_{X,Y} P(X,Y)\log  P(X,Y),$$
  is the joint Shannon entropy with the joint probability $P(X,Y)$ for both the variables $X$ and $Y$.

One can relate the joint probability to the conditional probability  as 
\begin{eqnarray}
P(X,Y)=P(X)P(Y|X)
\label{d2}
\end{eqnarray}
where $P(Y|X)$ is the probability of $Y$ if $X$ is given with probability $P(X)$. Thus the joint entropy $H(X,Y)$ can be rewritten as
\begin{eqnarray}
H(X,Y)=-\sum_{X,Y} P(X,Y)\left[\log  P(X)+\log P(Y|X) \right]
\label{d3}
\end{eqnarray}  
Hence we rewrite \ref{d1}  as
\begin{eqnarray}
{\cal I}(X,Y)=H(Y)-H(Y|X),
\label{d4}
\end{eqnarray}
where 
$$H(Y|X)=-\sum_{X,Y} P(X)P(Y|X)\log P(Y|X)),$$
 is regarded as the conditional entropy. It represents the average over $X$ of the Shannon entropy of $Y$, given $X$.

The above construction is purely classical. For a quantum system on the other hand, the Shannon entropy gets  replaced by the von Neumann entropy, $S(\rho)=-{\rm Tr}\rho\log \rho$,  where $\rho$ is the density operator. Also, $P(X,Y)$, $P(X)$ and $P(Y)$ get replaced  respectively by the  whole system density operator $\rho_{X,Y}$, the reduced density operator of the subsystem $X\,(\rho_{\small X}={\rm Tr}_{\footnotesize Y}\rho_{X,Y}$) and  the reduced density operator of the subsystem $Y\,(\rho_{Y}=\operatorname{Tr_{X}\rho_{X,Y}}$). 

The notion of the conditional probability $P(Y|X)$ in quantum mechanics requires projective measurements, achieved via a complete set of projection operators, $\Pi_{i}=|\psi_{i}\rangle\langle\psi_{i}|$, for all $i$. The density operator of $Y$ after we have measured $X$ is given by,
\begin{eqnarray}
\rho_{Y|i}=\frac{{\rm Tr}_{X}\left(\Pi_i\rho_{X,Y}\Pi_i \right)}{p_{i}}
\label{d5}
\end{eqnarray}
where $p_{i}=\operatorname{Tr}_{X,Y}(\Pi_{i}\rho_{X,Y}\Pi_{i})$. This allows us to define a quantum analogue of the conditional entropy,
\begin{eqnarray}
S(Y|X)=\min_{\Pi_{i}}\sum_{i}p_{i}S(\rho_{Y|i})
\label{d6}
\end{eqnarray}
where $S(\rho_{Y|i})$ is the von Neumann entropy for the density operator defined in \ref{d5}. The `min'
appearing in the above expression corresponds to measurements which disturb the system the least. This minimises the influence of the projectors. Putting these all in together, we write down the quantum analogues of \ref{d1}, \ref{d4}, respectively as
\begin{eqnarray}
{\cal I}_{X,Y}=S(\rho_{X})+S(\rho_{Y})-S(\rho_{X,Y}), \qquad 
{\cal J}_{X,Y}=S(\rho_{Y})-S(X|Y)
\label{d7}
\end{eqnarray}
However, it is clear that the classically equivalent expressions \ref{d1}, \ref{d4} of the mutual information need not necessarily be equal in the quantum scenario, for they involve different measurement procedure and quantum measurement  on one subsystem can  affect the other. 
Accordingly,  the quantum discord  is defined as~\cite{Zurek, Dakic},
\begin{eqnarray}
\mathscr{D}_{X,Y}:={\cal I}_{X,Y}-{\cal J}_{X,Y}=S(\rho_{X})-S(\rho_{X,Y})+S({Y|X}) 
\label{d8}
\end{eqnarray}
We wish to compute the above quantity in the cosmological de Sitter background we are interested in, for a maximally entangled `in' state,
\begin{eqnarray}
|\psi\rangle=\frac{1}{\sqrt2}(|0\rangle_{\vec{l}}|0\rangle_{\vec{k}}+|1\rangle_{\vec{l}}|1\rangle_{\vec{k}})
\label{d9}
\end{eqnarray}
Using the expression for the two-mode squeezed state \ref{sq2}, the above state can be expanded in terms of the `out' states. We shall consider two cases below. In the first case the states denoted by the momentum $\vec{l}$ will be held intact. In another case we shall consider their squeezed state expansion as well. This will help us to probe the correlations between the `in-out' and the `out-out' sectors.

\begin{figure}[h!]
\centering
  \includegraphics[width=6cm]{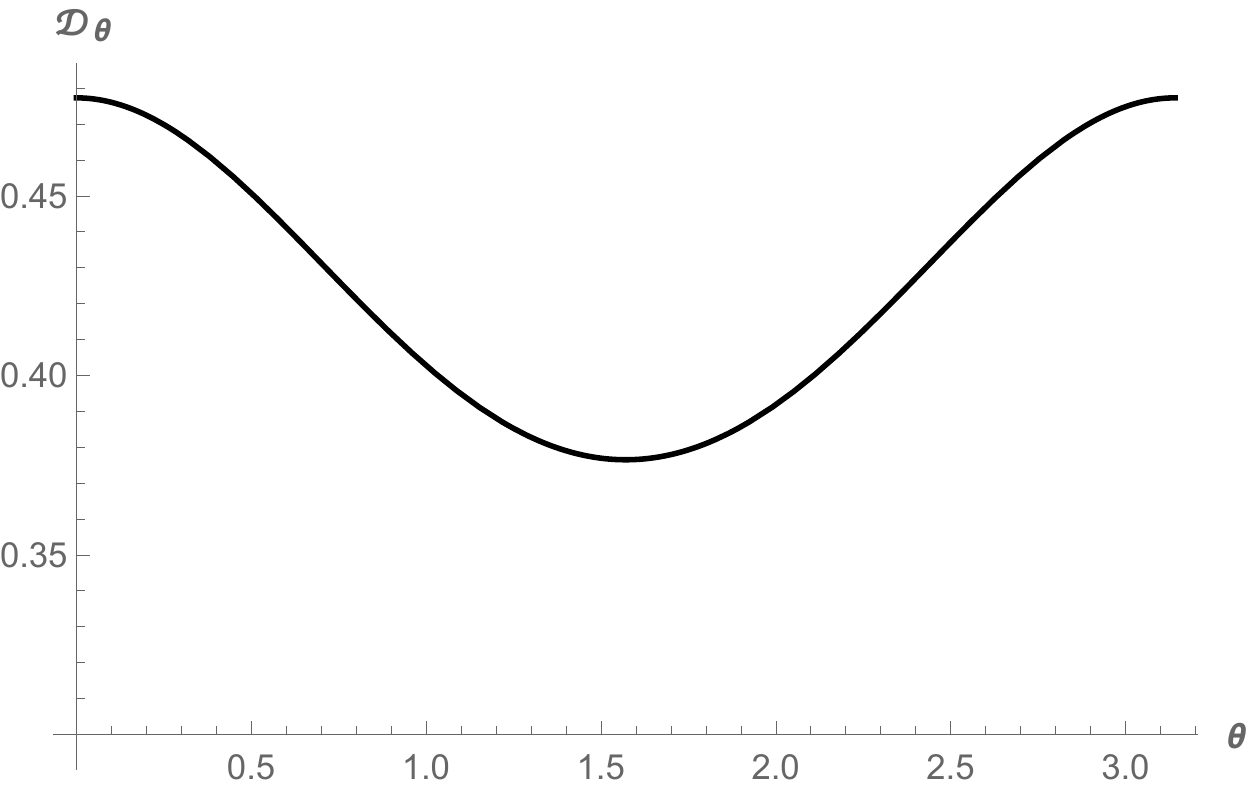} \hskip 1cm
  \includegraphics[width=7cm]{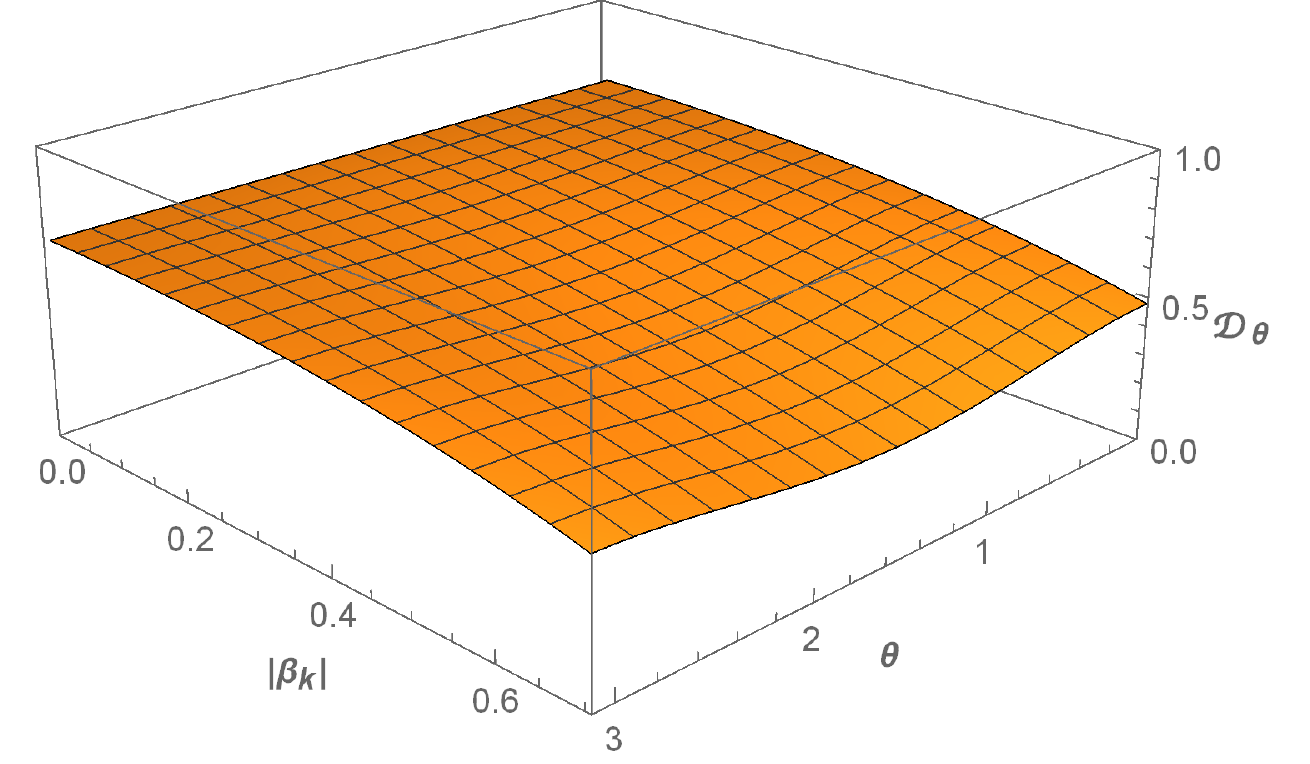}
 \caption{(Left) Plot for the quantum discord, \ref{d19}, with respect to the polar angle $\theta$ with $|\beta_k| \approx 0.707$. The minimisation occurs at $\theta =\pi/2$. This is valid for all possible values of the Bogoliubov coefficient. (Right) A 3-D plot for the same with the simultaneous variation of $\theta$ and $|\beta_{k}|$. As of \ref{bglv3}, the variation of the Bogoliubov coefficient correspond to the variation of the rest mass  only, and hence for a given species of field, $|\beta_k|$ or the quantum discord is fixed with respect to its spatial momentum ${\vec k}$.}
  \label{fig:discord1}
  \end{figure}

Accordingly as the first case, using \ref{sq2} for the modes $\vec{k}$, we rewrite \ref{d9} as
\begin{eqnarray}
|\psi\rangle=\frac{1}{\sqrt2}\frac{|\alpha_{k}|}{\alpha_{k}^\ast}\left[|0\rangle_{\vec{l}}\otimes\left(\alpha_{k}^\ast|0_{\vec{k},{\rm out}}\rangle |0_{-\vec{k},{\rm out}}\rangle - {\beta_{k}} |1_{\vec{k},{\rm out}}\rangle |1_{-\vec{k},{\rm out}}\rangle \right)+|1\rangle_{\vec{l}}\otimes (|1_{\vec{k},{\rm out}}\rangle|0_{-\vec{k},{\rm out}}\rangle)\right]
\label{d11}
\end{eqnarray}
The reduced density matrix $\rho_{XY}$ ($X \equiv \vec{l},\,\, Y \equiv \vec{k}$), after tracing out over $-\vec{k}$ and suppressing the level `out' for the sake of brevity is given by,
\begin{eqnarray}
\begin{split}
\rho_{XY}=&\frac{1}{2}\left[|0\rangle_{\vec{l}} {_{\vec{l}}\langle0|}\otimes(|\alpha_{k}|^2 |0\rangle_{\vec{k}} {_{\vec{k}}\langle0|})+|0\rangle_{\vec{l}} {_{\vec{l}}\langle0|}\otimes (|\beta_{k}|^2 |1\rangle_{\vec{k}} {_{\vec {k}}\langle1|})\right. \\&\left.+|1\rangle_{\vec{l}} {_{\vec{l}}\langle0|}\otimes (\alpha_{{k}} |1\rangle_{\vec{k}} {_{\vec{k}}\langle0|}) +|0\rangle_{\vec{l}} {_{\vec{l}}\langle1|}\otimes (\alpha^\ast_{k} |0\rangle_{\vec{k}} {_{\vec{k}}\langle1|})+|1\rangle_{\vec{l}} {_{\vec{l}}\langle1|}\otimes|1\rangle_{\vec{k}} {_{\vec{k}}\langle1|}\right]
\label{d12}
\end{split}
\end{eqnarray}
We also have,
\begin{eqnarray}
\begin{split}
\rho_{X}&=\operatorname{Tr}_{Y} (\rho_{X,Y})
=\frac{1}{2}[(|0\rangle_{\vec{l}} {_{\vec{l}}\langle0|}+|1\rangle_{\vec{l}} {_{\vec{l}}\langle1|})],\\
\rho_{Y}&=\operatorname{Tr}_{X} (\rho_{X,Y})
=\frac{1}{2}[|\alpha_{k}|^2(|0\rangle_{\vec{k}} {_{\vec{k}}\langle0|}+(1+|\beta_{k}|^2)|1\rangle_{\vec{k}} {_{\vec{k}}\langle1|})]\\&
\label{d13}
\end{split}
\end{eqnarray}
Using \ref{d12} and \ref{d13}, we now compute the von Neumann entropies,
\begin{eqnarray}
\begin{split}
&S(\rho_{X})=\log 2\\&
S(\rho_{Y})=-\frac{1}{2}\left[|\alpha_{k}|^2\log\frac{|\alpha_{k}|^2}{2}+(1+|\beta_{k}|^2)\log\frac{(1+|\beta_{k}|^2)}{2}\right]\\&
S(\rho_{XY})=-\frac{1}{2}\left[(1+|\alpha_{k}|^2)\log\frac{(1+|\alpha_{k}|^2)}{2}+(1-|\alpha_{k}|^2)\log\frac{(1-|\alpha_{k}|^2)}{2}\right]
\end{split}
\label{d14}
\end{eqnarray}
In order to compute the conditional entropy $S(Y|X)$, \ref{d6}, which needs a minimisation over the projective measurements, we take the usual projection operators~\cite{Zurek, Kanno:2016gas, A.dutta},
\begin{eqnarray}
\begin{split}
\Pi_{\pm}  :&= \frac{1}{2}\left[(1\pm{\hat x}_{3})|0\rangle_{\vec{l}} {_{\vec{l}}}\langle0|+(1\mp{\hat x}_{3})|1\rangle_{\vec{l}} {_{\vec{l}}}\langle1|\pm({\hat x}_{1}- i{\hat x}_{2})|0\rangle_{\vec{l}} {_{\vec{l}}}\langle1|\pm({\hat x}_{1}+i{\hat x}_{2})|1\rangle_{\vec{l}} {_{\vec{l}}}\langle0|\right]
%\frac{I \pm \hat{x}.{\sigma}}{2},\qquad  ,\\&=
%=\frac{1}{2} \begin{pmatrix} (1\pm{\hat x}_{3})  &   \pm({\hat x}_{1}-i {\hat x}_{2}) \\  \pm({\hat x}_{1}+i {\hat x}_{3})   &   (1\mp {\hat x}_{3}) \end{pmatrix}
\label{d15}
\end{split}
\end{eqnarray}
where $ {\hat x}_{1}^2+{\hat x}_{2}^2+{\hat x}_{3}^2=1$, representing spatial unit vectors.
Note that the above operates only on the $\vec{l}$ sector of \ref{d12}. Since the relevant  Hilbert space is 2-dimensional, we have taken two projectors which are orthogonal to each other and follows the identity $\Pi^2_{\pm}=\Pi_{\pm}$. 
Using  now $\operatorname{Tr}(AB)= \operatorname{Tr}(BA)$, we get from \ref{d5} after some algebra,
\begin{eqnarray}
\begin{split}
\rho_{Y|\pm}=&\frac{1}{2} \left[{(1\pm {\hat x}_{3})(|\alpha_{k}|^2 |0\rangle_{\vec{k}} {_{\vec{k}}\langle0|}+ |\beta_{k}|^2 |1\rangle_{\vec{k}} {_{\vec{k}}\langle1|})}+(1\mp {\hat x}_{3})(|1\rangle_{\vec{k}}{_{\vec{k}}}\langle1|)\right.\\&\left.\pm(\hat{x}_{1}-i {\hat x}_{2})(\alpha_{k}|1\rangle_{\vec{k}}{_{\vec{k}}}\langle0|)\pm({\hat x}_{1}+i {\hat x}_{2})(\alpha^\ast_{k}|0\rangle_{\vec{k}}{_{\vec{k}}}\langle1|)\right]
\end{split}
\label{d17}
\end{eqnarray}
In the usual parametrisation,
$${\hat x}_{1}=\sin{\theta} \cos{\phi},\qquad {\hat x}_{2}=\sin{\theta}\sin{\phi}, \qquad {\hat x}_{3}=\cos{\theta},$$
the conditional entropy is given by \ref{d6} and is found to be independent of the azimuthal angle $\phi$,
\begin{eqnarray}
\begin{split}
S(Y|X)&=-\frac{1}{2}\left[\left(\frac{1+\sqrt{1-(1+\cos{\theta})^2|\beta_{k}|^2|\alpha_{k}|^2}}{2}\right)\log\left(\frac{1+\sqrt{1-(1+\cos{\theta})^2|\beta_{k}|^2|\alpha_{k}|^2})}{2}\right)\right.\\&\left.+\left(\frac{1-\sqrt{1-(1+\cos{\theta})^2|\beta_{k}|^2|\alpha_{k}|^2})}{2}\right)\log\left(\frac{(1-\sqrt{1-(1+\cos{\theta})^2|\beta_{k}|^2|\alpha_{k}|^2})}{2}\right)\right.\\&\left.+\left(\frac{(1+\sqrt{1-(1-\cos{\theta})^2|\beta_{k}|^2|\alpha_{k}|^2})}{2}\right)\log\left(\frac{(1+\sqrt{1-(1-\cos{\theta})^2|\beta_{k}|^2|\alpha_{k}|^2})}{2}\right)\right.\\&\left.+\left(\frac{(1-\sqrt{1-(1-\cos{\theta})^2|\beta_{k}|^2|\alpha_{k}|^2})}{2}\right)\log\left(\frac{(1-\sqrt{1-(1-\cos{\theta})^2|\beta_{k}|^2|\alpha_{k}|^2})}{2}\right)\right]_{\rm min},
\end{split}
\label{d18}
\end{eqnarray}
where the suffix `min' stands for the minimisation with respect to $\theta$. The quantum discord, \ref{d8}, is given by,
\begin{eqnarray}
\begin{split}
{\mathscr{D}_{\theta}}
=&\log 2+\frac{1}{2}\left[(1+|\alpha_{k}|^2)\log\frac{(1+|\alpha_{k}|^2)}{2}+(1-|\alpha_{k}|^2)\log\frac{(1-|\alpha_{k}|^2)}{2}\right]\\&-\frac{1}{2}\left[\left(\frac{1+\sqrt{1-(1+\cos{\theta})^2|\beta_{k}|^2|\alpha_{k}|^2}}{2}\right)\log\left(\frac{1+\sqrt{1-(1+\cos{\theta})^2|\beta_{k}|^2|\alpha_{k}|^2})}{2}\right)\right.\\&\left.+\left(\frac{1-\sqrt{1-(1+\cos{\theta})^2|\beta_{k}|^2|\alpha_{k}|^2})}{2}\right)\log\left(\frac{(1-\sqrt{1-(1+\cos{\theta})^2|\beta_{k}|^2|\alpha_{k}|^2})}{2}\right)\right.\\&\left.+\left(\frac{(1+\sqrt{1-(1-\cos{\theta})^2|\beta_{k}|^2|\alpha_{k}|^2})}{2}\right)\log\left(\frac{(1+\sqrt{1-(1-\cos{\theta})^2|\beta_{k}|^2|\alpha_{k}|^2})}{2}\right)\right.\\&\left.+\left(\frac{(1-\sqrt{1-(1-\cos{\theta})^2|\beta_{k}|^2|\alpha_{k}|^2})}{2}\right)\log\left(\frac{(1-\sqrt{1-(1-\cos{\theta})^2|\beta_{k}|^2|\alpha_{k}|^2})}{2}\right)\right]_{\rm min}
\end{split}
\label{d19}
\end{eqnarray}
For different values of the Bogoliubov coefficients,  ${\mathscr{D}_{\theta}}$ appearing above minimises  at $\theta=\pi/2$,  \ref{fig:discord1}.  For the maximum value $|\beta_{k}|\approx 0.707$ in the cosmological de Sitter, we have  ${\mathscr{D}_{\pi/2}}\approx 0.377$. Note also that ${\mathscr{D}_{\theta}}$ is never vanishing, not even for $|\beta_k| \to 0$, in which limit it equals $\log 2$.
\\

\noindent
As we mentioned earlier,  we shall now examine the second scenario where  both $l$ and $k$ sectors in \ref{d9} undergo the squeezed state expansion, so that
\begin{eqnarray}
\begin{split}
|\psi\rangle&=\frac{|\alpha_{l}| |\alpha_{k}|}{\sqrt2 \alpha_{l}^\ast \alpha_{k}^\ast   }
\left[(\alpha_{l}^\ast |0_{\vec{l},{\rm out}}\rangle |0_{-\vec{l},{\rm out}}\rangle - {\beta_{l}} |1_{\vec{l},{\rm out}}\rangle |1_{-\vec{l},{\rm out}}\rangle)\otimes(\alpha_{k}^\ast|0_{\vec{k},{\rm out}}\rangle |0_{-\vec{k},{\rm out}}\rangle - {\beta_{k}} |1_{\vec{k},{\rm out}}\rangle |1_{-\vec{k},{\rm out}}\rangle)\right.+\\& \left. |1_{\vec{l},{\rm out}}\rangle |0_{-\vec{l},{\rm out}}\rangle \otimes |1_{\vec{k},{\rm out}}\rangle|0_{-\vec{k},{\rm out}}\rangle\right]
\end{split}
\label{d21}
\end{eqnarray}
We define the reduced density operator  $\rho_{XY}$ $(X\equiv \vec{l},\, Y\equiv \vec{k})$ by tracing out over the $(-\vec{l}, -\vec{k})$ sector,
\begin{eqnarray}
\begin{split}
\rho_{XY}&=\frac{1}{2}\left[(|\alpha_{l}|^2|0\rangle_{\vec{l}} {_{\vec{l}}\langle0|})\otimes(|\alpha_{k}|^2 |0\rangle_{\vec{k}} {_{\vec{k}}\langle0|})+(|\alpha_{l}|^2|0\rangle_{\vec{l}} {_{\vec{l}}\langle0|})\otimes (|\beta_{k}|^2 |1\rangle_{\vec{k}} {_{\vec {k}}\langle1|})\right.\\&\left.+(|\beta_{{l}}|^2|1\rangle_{\vec{l}} {_{\vec{l}}\langle1|})\otimes (|\alpha_{{k}}|^2 |0\rangle_{\vec{k}} {_{\vec{k}}\langle0|}) +(\alpha^\ast_{l}|0\rangle_{\vec{l}} {_{\vec{l}}\langle1|})\otimes (\alpha^\ast_{k} |0\rangle_{\vec{k}} {_{\vec{k}}\langle1|})\right.\\&\left.+(\alpha_{l}|1\rangle_{\vec{l}} {_{\vec{l}}\langle0|})\otimes (\alpha_{k} |1\rangle_{\vec{k}} {_{\vec{k}}\langle0|})+(|\beta_{l}|^2|1\rangle_{\vec{l}} {_{\vec{l}}\langle1|})\otimes(|\beta_{k}|^2|1\rangle_{\vec{k}} {_{\vec{k}}\langle1|})+|1\rangle_{\vec{l}} {_{\vec{l}}\langle1|}\otimes|1\rangle_{\vec{k}} {_{\vec{k}}\langle1|}\right]
\label{d22}
\end{split}
\end{eqnarray}
where we have suppressed the level `out' as earlier for the sake of brevity. We find after some algebra, the following von Neumann entropies,
\begin{eqnarray}
\begin{split}
&S(\rho_{X})=-\frac{1}{2}\left[|\alpha_{l}|^2\log\frac{|\alpha_{l}|^2}{2}+(1+|\beta_{l}|^2)\log\frac{(1+|\beta_{l}|^2)}{2}\right]\\&
S(\rho_{Y})=-\frac{1}{2}\left[|\alpha_{k}|^2\log\frac{|\alpha_{k}|^2}{2}+(1+|\beta_{k}|^2)\log\frac{(1+|\beta_{k}|^2)}{2}\right]\\&
S(\rho_{XY})=-\frac{1}{2}\left[|\beta_{l}|^2|\alpha_{k}|^2\log\frac{|\beta_{l}|^2|\alpha_{k}|^2}{2}+|\beta_{k}|^2|\alpha_{l}|^2\log\frac{|\beta_{k}|^2|\alpha_{l}|^2}{2}\right.\\&\left.+\left(\frac{(1+|\alpha_{l}|^2|\alpha_{k}|^2+|\beta_{l}|^2|\beta_{k}|^2)+\sqrt{(1+|\alpha_{l}|^2|\alpha_{k}|^2+|\beta_{l}|^2|\beta_{k}|^2)^2-4|\alpha_{l}|^2|\alpha_{k}|^2|\beta_{l}|^2|\beta_{k}|^2}}{2}\right)\right.\\&\left. \times \log\left(\frac{(1+|\alpha_{l}|^2|\alpha_{k}|^2+|\beta_{l}|^2|\beta_{k}|^2)+\sqrt{(1+|\alpha_{l}|^2|\alpha_{k}|^2+|\beta_{l}|^2|\beta_{k}|^2)^2-4|\alpha_{l}|^2|\alpha_{k}|^2|\beta_{s}|^2|\beta_{k}|^2}}{4}\right)\right.\\&\left.+\left(\frac{(1+|\alpha_{l}|^2|\alpha_{k}|^2+|\beta_{l}|^2|\beta_{k}|^2)-\sqrt{(1+|\alpha_{l}|^2|\alpha_{k}|^2+|\beta_{l}|^2|\beta_{k}|^2)^2-4|\alpha_{l}|^2|\alpha_{k}|^2|\beta_{l}|^2|\beta_{k}|^2}}{2}\right)\right. \\&\left.\times \log\left(\frac{(1+|\alpha_{l}|^2|\alpha_{k}|^2+|\beta_{l}|^2|\beta_{k}|^2)-\sqrt{(1+|\alpha_{l}|^2|\alpha_{k}|^2+|\beta_{l}|^2|\beta_{k}|^2)^2-4|\alpha_{l}|^2|\alpha_{k}|^2|\beta_{l}|^2|\beta_{k}|^2}}{4}\right)\right]
\end{split}
\label{d23}
\end{eqnarray}
 Since the $\vec{l}$ sector of \ref{d22} is two dimensional by the virtue of the Pauli exclusion principle, the conditional entropy, \ref{d6}, for this system can still be computed as earlier using the projectors of~\ref{d15}.\footnote{For Bosons on the other hand, we would have obtained an infinite dimensional Hilbert space for the $\vec{l}$ sector, requiring the construction of an infinite number of projection operators to define the quantum discord.  One possible way to tackle this issue seems to be the   truncation of the relevant squeezed state expansion at some finite order by assuming  $|\beta_l|/|\alpha_l| \ll 1$. A full resolution of this situation, however, is not clear to us.} We find
\begin{eqnarray}
\begin{split}
S({Y|X})&=p_{+}S(\rho_{Y|+})+p_{-}S(\rho_{Y|-})\\&
=p_{+}\left[-\left(\frac{2p_{+}+\sqrt{(2p_{+})^2-C_{+}}}{4p_{+}}\right)\log\left(\frac{2p_{+}+\sqrt{(2p_{+})^2-C_{+}}}{4p_{+}}\right)\right.\\&\left.-\left(\frac{2p_{+}-\sqrt{(2p_{+})^2-C_{+}}}{4p_{+}}\right)\log\left(\frac{2p_{+}-\sqrt{(2p_{+})^2-C_{+}}}{4p_{+}}\right)\right] \\&+p_{-}\left[-\left(\frac{2p_{-}+\sqrt{(2p_{-})^2-C_{-}}}{4p_{-}}\right)\log\left(\frac{2p_{-}+\sqrt{(2p_{-})^2-C_{-}}}{4p_{-}}\right)\right.\\&\left.-\left(\frac{2p_{-}-\sqrt{(2p_{-})^2-C_{-}}}{4p_{-}}\right)\log\left(\frac{2p_{-}-\sqrt{(2p_{-})^2-C_{-}}}{4p_{-}}\right)\right]_{\rm min}
\end{split}
\label{d28}
\end{eqnarray}
where $p_{\pm}$ and  $C_{\pm}$ are given by
\begin{eqnarray}
\begin{split}
{p_{\pm}}&=\operatorname{Tr}_{X,Y}\left(\Pi_{\pm}\rho_{X,Y}\Pi_{\pm}\right)=\frac{1}{2}\left(1\mp \cos\theta|\beta_{k}|^2\right),   \\
C_{\pm}&={(1\pm \cos\theta)^2|\alpha_{l}|^4|\alpha_{k}|^2|\beta_{k}|^2}+2 |\beta_{l}|^2|\beta_{k}|^2|\alpha_{l}|^2|\alpha_{k}|^2\,\sin^2\theta+(1\mp\cos\theta)^2 |\beta_{l}|^2|\alpha_{k}|^2(|\beta_{l}|^2|\beta_{k}|^2+1)
\end{split}
\label{d29}
\end{eqnarray}
and the suffix `min' in \ref{d28} refers to minimisation with respect to the polar angle $\theta$. Note that alike \ref{d18}, the above is independent of the azimuthal angle $\phi$.

\begin{figure}[h!]
\centering
  \includegraphics[width=6cm]{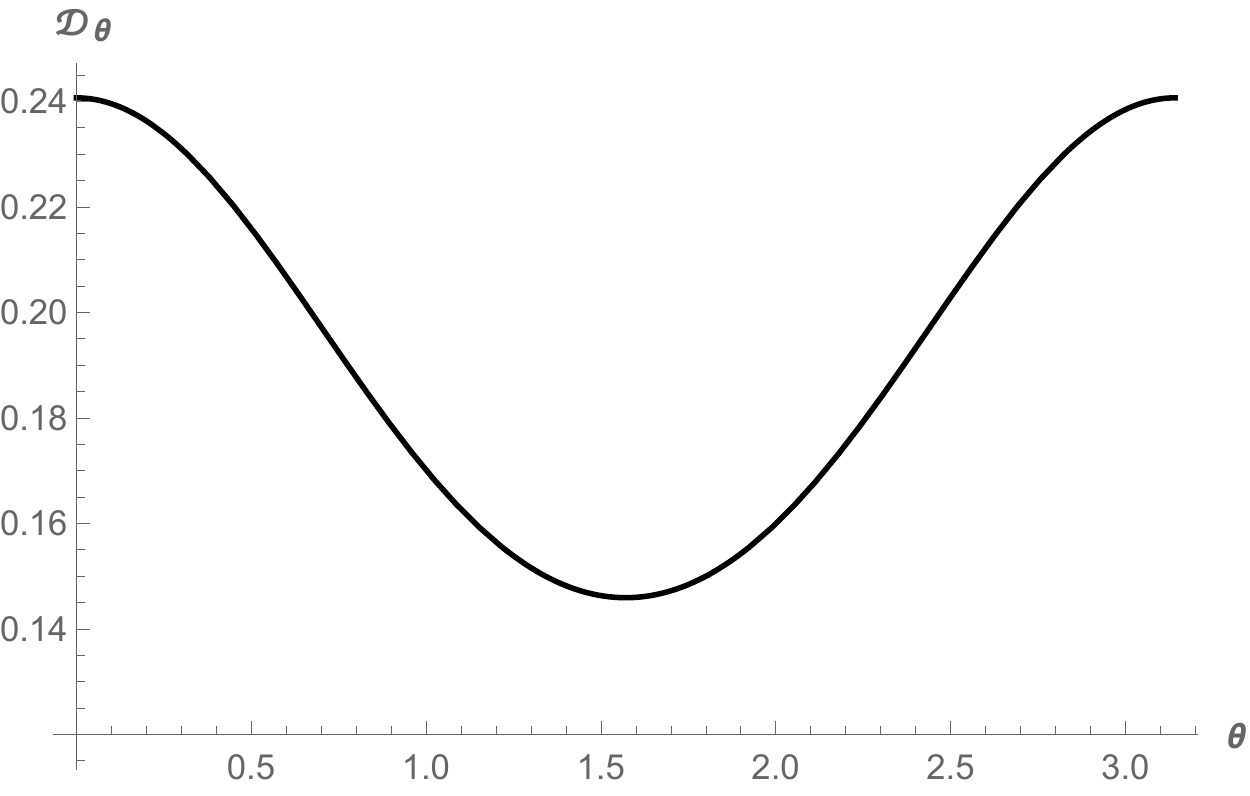} \hskip 1cm
   \includegraphics[width=7cm]{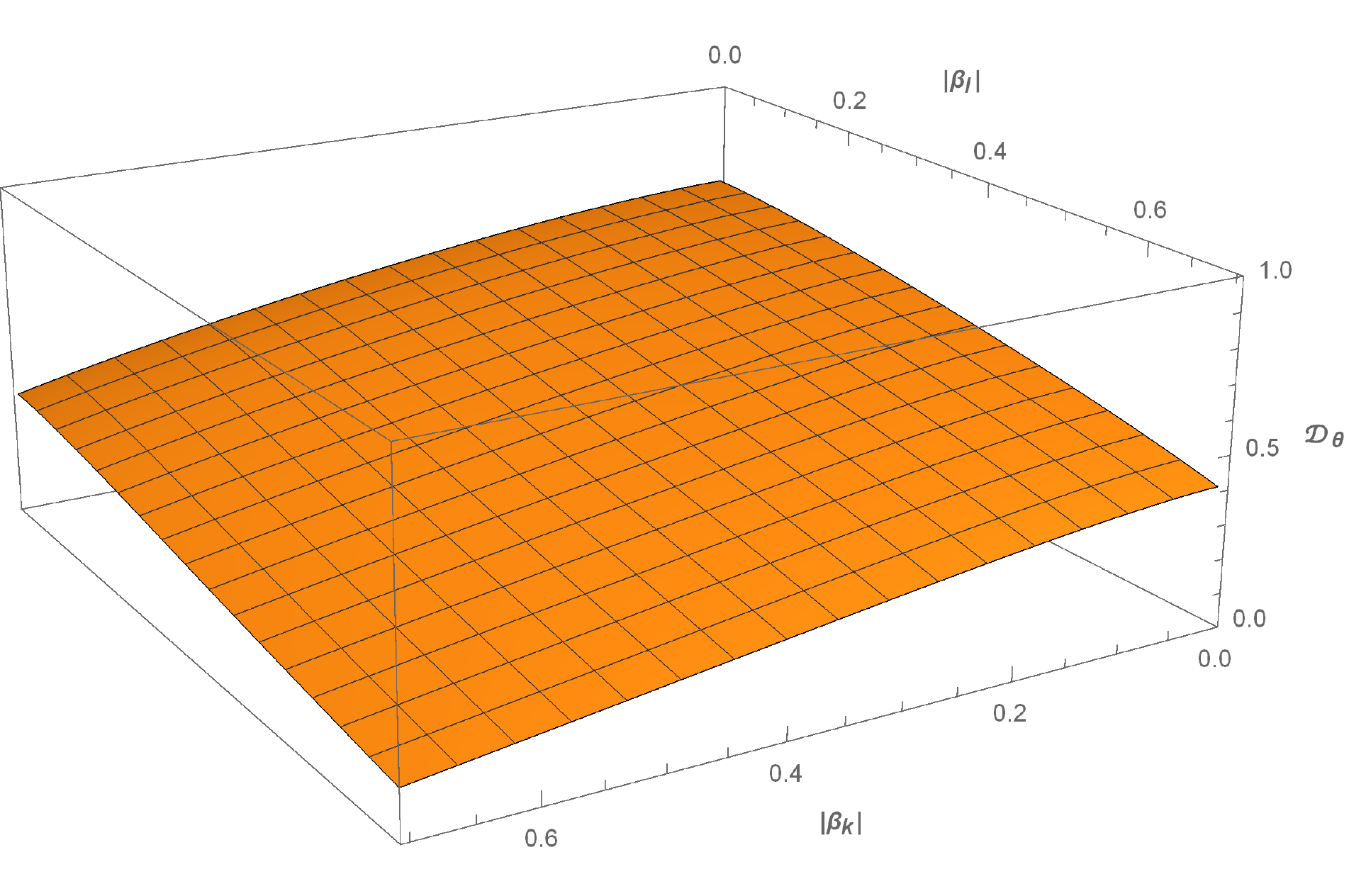}
  \caption{(Left) Plot for the quantum discord characterising `out-out' correlations, with respect to the polar angle $\theta$ with $|\beta_l|=|\beta_k| \approx 0.707$. The minima occurs at $\theta =\pi/2$ for all possible values of the Bogoliubov coefficient, similar to the `in-out' sector, \ref{fig:discord1}. (Right) A 3-D plot of the same, depicting its variation with respect to the two Bogoliubov coefficients, $|\beta_{k}|$ and $|\beta_{l}|$. As of \ref{bglv3}, the variation of the Bogoliubov coefficients correspond to the variation of the rest mass only.}
  \label{fig:discord}
\end{figure}

Even though the $\theta$ dependence of \ref{d29} looks  different compared to \ref{d18}, the discord, ${\mathscr{D}_{\theta}}=S(\rho_{X})-S(\rho_{X,Y})+S({Y|X})$, still minimises at $\theta = \pi/2$, as depicted in \ref{fig:discord}. For the Bogoliubov coefficients close to their maximal values, $|\beta_{l}|=|\beta_{k}|\approx 0.707$, we have ${\mathscr{D}_{\pi/2}}\approx 0.146$, which is less compared to our previous case, \ref{fig:discord1}. This shows the degradation of correlations in the `out-out' states compared to the `in-out' states. We also have plotted the variation of the discord with respect to the two Bogoliubov coefficients in \ref{fig:discord}. As earlier, the discord is never vanishing.
%
%\begin{figure}[h!]
%\begin{center}
  %\includegraphics[width=7cm]{f7.pdf}
  %\caption{3D plot of the quantum discord characterising `out-out' correlations, with the variation of the two Bogoliubov coefficients, $|\beta_{k}|$ and $|\beta_{l}|$. As of \ref{bglv3}, the variation of the Bogoliubov coefficients correspond to the variation of the rest mass only.}
  %\label{fig:discord'}
%\end{center}
%\end{figure}
%

The above analysis was done by tracing out the $(-\vec{k}, -\vec{l})$ subsectors in \ref{d21}.  One can also trace out the other subsectors and analogously compute the quantum discord between them. We expect qualitatively similar results to hold. We also note the qualitative similarity of \ref{fig:discord1}, \ref{fig:discord} with the earlier studies on scalar and fermion fields in the Rindler spacetime~\cite{Brown:2012iz, A.dutta, wang} and as well as on the scalar field in the hyperbolic de Sitter spacetime~\cite{Kanno:2016gas}.

Finally, we wish to mention very briefly the logarithmic negativity~\cite{Zyczkowski:1998yd, Vidal:2002zz} for this system. The log-negativity is a measure of pure quantum correlations,  useful in particular, for mixed ensembles. In order to compute this, one needs to take the partial transpose of the matrix representation of \ref{d12},
%
%\begin{eqnarray}
$$(\rho_{XY})^{\rm T}=\frac{1}{2} \begin{pmatrix} |\alpha_{k}|^2 & 0 & 0 & 0 \\   0 & |\beta_{k}|^2 &\alpha_{k}^\ast & 0 \\ 0 &\alpha_{k} & 0 & 0 \\ 0 & 0& 0&1  \end{pmatrix},$$
%\label{d30}
%\end{eqnarray}
%
whose eigenvalues are $1/2,\,1/2,\, \pm |\alpha_{k}|^2/2$. The log-negativity is given by,
\begin{eqnarray}
{\cal{L}_N}=\log\left({1+{|\alpha_{k}|^2}}\right)
\label{d31}
\end{eqnarray}
Since we do not have any extreme squeezing limit $(|\beta_k| \to 1,\,|\alpha_k| \to 0)$ for our case, cf. \ref{A}, the above quantity is always non-vanishing.  Similar conclusion holds for the mixed density operator of \ref{d22} as well. This is qualitatively different for the scalar field theory  in the context of transition from the de Sitter to radiation dominated era~\cite{Kanno:2016gas}, where such extreme squeezing may happen for which  
the log-negativity vanishes, even though the quantum discord remains non-vanishing. For our case {\it both} log-negativity and the quantum discord are always non-vanishing.

%%%%%%
\section{Discussions}\label{con}
%%%%%%

In this work we have computed the BMK violation and quantum discord for massive Dirac fermions in the cosmological de Sitter background, respectively in \ref{Bell} and \ref{discord}. For the BMK violation, we have focussed on the vacuum corresponding to the two- and four-mode squeezed states whereas for the latter we have used a maximally entangled Bell state as our `in' state. Our motivation was to see how the results differ subject to different coordinatisation of the de Sitter as well as the spin of the field. We thus first note the qualitative similarities of the variations with respect to parametrisation angles  (\ref{fig:bell1}, \ref{fig:bell3}, \ref{fig:discord1}, \ref{fig:discord}) with a scalar field theory discussed respectively in the context of transition from de Sitter to radiation dominated era~\cite{Kanno:2017dci} and the hyperbolic de Sitter spacetime~\cite{Kanno:2016gas}. However, the chief qualitative difference for the cosmological de Sitter from them or the non-inertial frame (e.g.~\cite{Alsing:2003es}), or even the static de Sitter spacetime (e.g.~\cite{Bhattacharya:2019zno}) is that the Bogoliubov coefficients for the `in' and `out' vacua in this case is independent of the spatial momentum or the total energy, \ref{bglv3}. Similar feature is seen for a scalar field theory in the global or the cosmological de Sitter spacetime, e.g.~\cite{Mottola:1984ar, Markkanen:2016aes}.  Thus as we have indicated earlier (e.g.~\ref{fig:discord1}), the variation of the Bogoliubov coefficients correspond to the variation of the rest mass only. In other words, unlike the standard R-L entanglement, for a given particle species, any of our results in the cosmological de Sitter spacetime  is fixed. 

We also note that there exists no extreme squeezing limit ($|\beta_k| \to 1$) for the states constructed in \ref{sqz} and we always have $|\beta_k| \lesssim 0.707$, as discussed at the end of \ref{A}. Due to this reason, as discussed at the end of the preceding section, the logarithmic negativity is never vanishing.  This is qualitatively different from the 
scenario reported in~\cite{Kanno:2016gas} for a scalar field theory where the log-negativity can indeed vanish, indicating the complete decay of quantum entanglement due to particle creation but the discord, being a measure of all correlations,  survive.   

Investigation of these results and also the decoherence properties in the presence of background primordial electric and magnetic fields seems interesting, due to the Schwinger pair creation mechanism. We hope to return to this issue in a future work.

\bigskip
%%%%%%%%%%%%%%%%%%%%%%%%%
\section*{Acknowledgement}
The research of SB is partially supported by the ISIRD grant 9-289/2017/IITRPR/704. Majority of HG's work was done when he was a Masters student at Indian Institute of Technology Ropar, India, and he also  acknowledges the same grant for partial support. HG's current research is supported   under the CSIR SPMF scheme of CSIR, India.

\bigskip
\appendix
\labelformat{section}{Appendix #1} 
%%%%%%%%%%%%%%%%%%%%%%%%%%%%%%%%%%%%%%%%%%%%%%%%%%%%
\section{The Dirac mode functions and Bogoliubov coefficients}\label{A}
%%%%%%%%%%%%%%%%%%%%%%%%%%%%%%%

In this Appendix we shall briefly review the solutions of the Dirac equation in the cosmological de Sitter background. The Greek indices appearing below will stand for the spacetime whereas the Latin ones will stand for the local Lorentz frame.

The de Sitter metric in the conformally flat form reads, 
\begin{eqnarray}
ds^2 = \frac{1}{H^2 \eta^2}(-d\eta^2 +dx^2+dy^2+dz^2)
\label{c0}
\end{eqnarray}
where $H= \sqrt{\Lambda/3}$ with $\Lambda$ being the cosmological constant, and  the conformal time $\eta$ varies from, $ -\infty < \eta <0$. We shall also require for our purpose the standard cosmological time $t$, given by
$$ t = -\frac{1}{H} \ln  (-H \eta), \qquad -\infty < t< \infty $$
The Dirac equation  in a curved background reads,
\begin{eqnarray}
\left[ i\gamma^\mu D_{\mu}-m\right] \Psi =0,
\label{c1}
\end{eqnarray}
where $D_{\mu}$ is the spin covariant derivative e.g.~\cite{Parker:2009uva},
	\begin{eqnarray}
	D_\mu:=\partial_\mu+\frac{1}{2}\omega_{\mu a b}\Sigma^{ab},
	\end{eqnarray}
where $\Sigma^{ab}=\left[\gamma^a,~\gamma^b\right]/4$ and the Ricci rotation coefficients $\omega$'s are given by 
 $$\omega_\mu{}^a{}_b = - e_b{}^\nu \left(\partial_\mu e^a{}_\nu - \Gamma^\lambda_{\mu \nu} e^a{}_\lambda \right)$$
The Gamma matrices appearing in \ref{c1} can be expanded in terms of the tetrad as, $\gamma^{\mu}= e^{\mu}{}_a \gamma^a$ and they satisfy the anti-commutation relation
$$
\left[\gamma^\mu, \gamma^\nu\right]_+ = 2g^{\mu\nu}\, {\bf I_{4\times 4}} 
$$
Making the choice of the tetrad,
\begin{eqnarray}
e^{\mu}{}_a  \equiv H\eta \, {\rm diag}\, \left(1, \, 1,\, 1,\, 1 \right),
\label{c2}
\end{eqnarray}
 the Dirac equation in the de Sitter  background becomes,  
\begin{eqnarray}
\left[i\gamma^0\eta \partial_{\eta} -\frac{3i \gamma^0}{2}+i\eta\, {\vec \gamma}\cdot{\vec \partial} -\frac{m}{H}\right]\Psi=0,
\end{eqnarray}

$\Psi(x)$ can be quantised  in terms of the conformal time as,
\begin{eqnarray}
\Psi(x)= \int \frac{d^{3} \vec{ k}}{(2\pi)^{3/2}}\,  \sum_{s= 1}^2 \left[a_{\rm in}(\vec{k},s) u_{\vec{k}, {\rm in }}^{(s)}(\eta) e^{i {\vec{k}\cdot \vec{x}}} + b_{\rm in}^{\dagger}(\vec{k},s) v_{\vec{k}, {\rm in }}^{(s)}(\eta)e^{-i \vec{k}\cdot \vec{x}} \right],
\label{f1}
\end{eqnarray}
where the temporal part of the Bunch-Davies mode functions are given by, e.g.~\cite{Collins:2004wj},
\begin{eqnarray}
\begin{split}
u_{\vec{k}, {\rm in }}^{(s)}(\eta)=\frac{\sqrt{\pi k}\, e^{\frac{m\pi}{2H}}}{2} \, \eta^2  \begin{pmatrix} H_{\nu}^{(2)}(k\eta) \\ i \beta_s  H_{\nu-1}^{(2)}(k\eta) \end{pmatrix} \phi_{ \hat{{k}}}^{(s)}, \qquad
v_{\vec{k}, {\rm in }}^{(s)}(\eta)=\frac{ \sqrt{\pi k}\, e^{\frac{m\pi}{2H}}}{2}\,  \eta^2  \begin{pmatrix} \beta_s H_{\nu}^{(1)}(k\eta) \\ -i   H_{1-\nu}^{(1)}(k\eta) \end{pmatrix} \chi_{ \hat{k}}^{(s)},
\label{inmodes}
\end{split}
\end{eqnarray}
where $\beta_s=\pm1$ is the helicity of the mode corresponding respectively to $s=1,2$. $\phi_{ \hat{k}}^{(s)}$ and $\chi_{\hat{k}}^{(s)}$ are eigenvectors of the helicity operator $\hat{k}\cdot {\vec\gamma}$, with eigenvalues $\beta_s$. They are given by, 
\begin{eqnarray}
\begin{split}
 &\phi_{\hat{k} }^{(1)}=- \chi_{\hat{k} }^{(2)}=\frac{e^{-i\delta}}{\sqrt 2} \begin{pmatrix} \frac{{\hat{k}_{x}}-i{\hat{k}_y}}{\sqrt{1-{\hat{k}_{x}}}} \\  \sqrt{1-{\hat{k}_{z}}} \end{pmatrix}, \qquad
\phi_{\hat{k} }^{(2)}= \chi_{\hat{k} }^{(1)}=\frac{e^{i\delta}}{\sqrt 2} \begin{pmatrix} -\sqrt{1-{\hat{k}_{z}}} \\   \frac{{\hat{k}_{x}}+i{\hat{k}_{y}}}{\sqrt{1-{\hat{k}_{x}}}} \end{pmatrix}\\&
\qquad \qquad \quad \quad  \phi_{-\hat{ {k}} }^{(s)}= \chi_{\hat{k} }^{(s)}, \qquad  e^{2i\delta}=- \frac {\left(\hat{k}_y+i \hat{k}_x \right)}{\sqrt{\hat{k}_x^2+\hat{k}_y^2}}   \label{spinors}
\end{split}
\end{eqnarray}
where $\hat{k}_x , \,\hat{k}_y$ are  momentum unit vectors.  Also, $\nu$ is a constant given by
$$\nu=\frac12+ \frac{im}{H}$$
$H_{\nu}^{(1)}(k\eta)$, $H_{\nu}^{(2)}(k\eta)$ are the Hankel functions of the first and second kind respectively. 
It is easy to check using their asymptotic properties~\cite{AS} that $u(\eta \to -\infty)\sim e^{-ik\eta}$ and $v(\eta \to -\infty) \sim e^{ik\eta}$ in \ref{inmodes} and thus respectively behave as positive and negative frequency solutions in the asymptotic past. The orthonormality of the mode functions appearing in \ref{f1} can be checked using \ref{spinors} by computing their inner products, say on an $\eta \to -\infty$ hypersurafce. 
\ref{inmodes} should be regarded as the fermionic analogue of the Bunch-Davies mode functions~\cite{Collins:2004wj, Bunch:1978yq, Spindel}.

The `out' modes are those which have definite positive and negative frequency characteristics in the asymptotic future, $\eta\to 0^-$. It is easy to see from the asymptotic expansion of the Hankel function  that the conformal time cannot be a good coordinate for  this purpose and accordingly we  resort to the cosmological time, $t$.  In terms of $t$,  we have from \ref{inmodes} in the asymptotic future, 
\begin{eqnarray}
\begin{split}
u_{\vec{k}, {\rm in }}^{(s)}(t\to \infty)&=&\frac{\sqrt{\pi k} \,e^{\frac{m\pi}{2 H}}}{2\pi H^2}  e^{-\frac{3Ht}{2}}  \begin{pmatrix} \Gamma(\nu) \left(-\frac{k}{2H}\right)^{-\nu}\,e^{iHmt} \\  \beta_s e^{-\frac{m\pi}{H}}\Gamma{(1-\nu)} \left({-\frac{k}{2H}}\right)^{\nu-1}e^{-iHmt}\end{pmatrix} \phi_{\hat{k} }^{(s)}
\\
v_{\vec{k}, {\rm in }}^{(s)}(t \to \infty)&=&\frac{\sqrt{\pi k} \,e^{\frac{m\pi}{2H}}}{2\pi H^2}  e^{-\frac{3Ht}{2}}  \begin{pmatrix} - \beta_s e^{-\frac{m\pi}{H}}\Gamma(\nu) \,\left({-\frac{k}{2H}}\right)^{-\nu}e^{iHmt} \\  \Gamma{(1-\nu)} \left({-\frac{k}{2H}}\right)^{\nu-1}e^{-iHmt}\end{pmatrix} \chi_{\hat{k}}^{(s)}
\label{out1}
\end{split}
\end{eqnarray}
Accordingly, we choose the temporal part of the `out' modes  as,
\begin{eqnarray}
\begin{split}
&u_{\vec{k}, {\rm out }}^{(s)}(t)= e^{-\frac{3Ht}{2}}\, e^{iHmt} \begin{pmatrix} 1 \\  0  \end{pmatrix} \phi_{\hat{k} }^{(s)}, \qquad
v_{\vec{k}, {\rm out }}^{(s)}(t)={e^{-\frac{3Ht}{2}}}  e^{-iHmt}\begin{pmatrix} 0 \\ 1  \end{pmatrix} \chi_{\hat{k} }^{(s)}
\label{out2}
\end{split}
\end{eqnarray}
Thus we have the field quantisation,
\begin{eqnarray}
\Psi(x)= \int \frac{d^{3} \vec{k}}{(2\pi)^{3/2}}\,  \sum_{s= 1}^2 \left[a_{\rm out}(\vec{k},s) u_{\vec{k}, {\rm out }}^{(s)}(t) e^{i {\vec{k}\cdot  \vec{x}}} + b_{\rm out}^{\dagger}(\vec{k},s) v_{\vec{k}, {\rm out }}^{(s)}(t)e^{-i \vec{k}\cdot \vec{ x}} \right],
\label{f2}
\end{eqnarray}

Equating \ref{f1} and \ref{f2} and using \ref{spinors}, \ref{out1}, we find the following Bogoliubov relations, 
\begin{eqnarray}
\begin{split}
a_{\rm out}(\vec{k},s)&=&\frac{e^{\frac{m \pi}{2H}}}{\sqrt{2\cosh \frac{m \pi }{H}}} \left[a_{\rm in}(\vec{k},s) -e^{-\frac{m\pi}{H}} b^{\dagger}_{\rm in}(-\vec{k},s)\right],\\
b^{\dagger}_{\rm out}(\vec{k},s)&=&\frac{e^{\frac{m \pi}{2H}}}{\sqrt{2\cosh \frac{m \pi }{H}}}  \left[b^{\dagger}_{\rm in}(\vec{k},s) +e^{-\frac{m\pi}{H}} a_{\rm in}(-\vec{k},s)\right]
\label{bglv3}
\end{split}
\end{eqnarray}
For any such  $(\vec{k}, s)$ mode we have the fermionic black body distribution with temperature $H/2\pi$,
\begin{eqnarray}
\langle0_{\rm in}|a^{\dagger}_{\rm out} a_{\rm out}|0_{\rm in}\rangle=|\beta_{{k}}|^2 =\frac1 {e^{\frac{2\pi m}{H}}+1}
\label{crp}
\end{eqnarray}
where 
$$|\alpha_{k}| =\frac{e^{\frac{m \pi}{2H}}}{\sqrt{2\cosh \frac{m \pi }{H}}}, \qquad |\beta_k|=  \frac{e^{-\frac{m \pi}{2H}}}{\sqrt{2\cosh \frac{m \pi }{H}}} \qquad ({\rm for~ both~ }s=1,2)$$
where the absolute values indicate that the Bogoliubov relations can be defined up to some global phase factors.
Note that \ref{crp} is independent of $k$. This originates from the fact that the frequencies of the `out' modes, \ref{out2} are solely determined by the rest mass $m$ of the field. Similar thing occurs for  scalar field theory in the global and the cosmological de Sitter backgrounds~\cite{Mottola:1984ar, Markkanen:2016aes}. Although we are certain that this result for fermions must have been reported earlier, we were unable to locate any particular reference stating the same.

 Recall also that a conformally invariant field like a massless fermion cannot create particles in the conformal vacuum of a conformally flat spacetime such as the de Sitter, e.g.~\cite{Parker:2009uva}. Thus we have an apparent ambiguity with setting $m= 0$ in \ref{crp}. However, we note that in this case the modes \ref{out2} have no positive and negative frequency characteristics at all. Thus it seems that the $m\to 0$ limit for \ref{crp}  is not smooth and one needs to treat the purely massless case separately, retaining explicitly its conformal symmetry by using the conformal vacuum. However, this case will not be relevant for our current purpose. Finally, we note the range of the Bogoliubov coefficient,
$$ 0 \leq |\beta_k| \lesssim 0.707$$
where the upper bound correspond to $m/H \to 0$.

%%%%%%%%%%%%%%%%%%%%%%%%

\end{document}